\documentclass[aps,prd,twocolumn,amsmath,showpacs,floatfix,amssymb,superscriptaddress,nofootinbib]{revtex4-1}
\newlength{\picwidth}

\makeatother
\usepackage{graphicx,epstopdf}
\usepackage{dcolumn}
\usepackage{bm}
\usepackage{amssymb,amsmath}
\usepackage{latexsym}
\usepackage{color}
\usepackage[normalem]{ulem}
\usepackage{cancel}
\allowdisplaybreaks
\usepackage{amsfonts}
\usepackage{color}
\usepackage{subfigure}
\newcommand{\be}{\begin{equation}}
\newcommand{\ee}{\end{equation}}
\newcommand{\f}{\frac}

\def\bfq{{\bf q}}

\def\bfk{{\bf k}}
\def\bfn{{\bf n}}

\def\bfzero{{\bf 0}}

\def\bfomega{\mbox{\boldmath $\omega$}}

\graphicspath{%
     {  }
}

\newcommand{\cM}[1]{\textcolor{magenta}{ #1}}

\newcommand{\bma}{\begin{pmatrix}}
\newcommand{\ema}{\end{pmatrix}}

\newcommand{\stell}{\affiliation{ Physics Department, Stellenbosch University,  
7602, South Africa}}
\newcommand{\stellb}{\affiliation{ Department of Applied Mathematics, Stellenbosch University,  
7602, South Africa}}
\newcommand{\nithep}{\affiliation{National Institute for Theoretical Physics (NITheP), Bag X1 Matieland, Stellenbosch, 7602, South Africa}}
\newcommand{\Maryland}{\affiliation{Maryland Center for Fundamental Physics \& Joint Space-Science Institute, Department of Physics, University of Maryland, College Park, MD 20742, USA}}
\newcommand{\OX}{\affiliation{Department of Astrophysics, University of Oxford, UK}}

\begin{document}

\title{The Astrophysics of Resonant Orbits in the Kerr Metric}

\author{Jeandrew Brink} \nithep \stellb
\author{Marisa Geyer} \stell \OX
\author{Tanja Hinderer} \Maryland 

\date{\today}

\begin{abstract}
This paper gives a complete
characterization of the location of resonant orbits in a Kerr spacetime for all possible
black hole spins and orbital parameter values.  A resonant orbit in this work is defined
as a geodesic for which the longitudinal and radial orbital frequencies are commensurate. 
Our analysis is based on expressing the resonance condition in its most transparent form using Carlson's elliptic integrals, which enable us to provide exact results together with a number of concise formulas characterizing the explicit dependence on the system parameters.
The locations of resonant orbits identify regions where intriguing observable phenomena could occur in astrophysical situations where various sources of perturbation act on the binary system . Resonant effects may have observable implications for the in-spirals of compact objects into a
super-massive black hole. During a generic in-spiral the slowly evolving
orbital frequencies will pass through a series of low-order  resonances where the ratio of orbital frequencies is equal to the ratio of two small integers. 
At these locations rapid changes in the
orbital parameters could  produce a measurable phase shift in the emitted gravitational and electromagnetic radiation. Resonant
orbits may also capture  gas or larger objects leading to
further observable characteristic electromagnetic emission. According to the KAM theorem, low order resonant orbits demarcate
the regions where the onset of geodesic chaos could occur when the Kerr Hamiltonian is perturbed.
 Perturbations are induced for example if the spacetime of the central object is non-Kerr, if gravity is modified, if the orbiting particle has large multipole moments, or if additional masses are nearby. We find
that the $1/2$ and $2/3$ resonances occur at  approximately
$4$ and $5.4$ Schwarzschild radii ($R_s$) from the black hole's event horizon. For compact object
in-spirals into super-massive black holes ($\sim 10^6 M_\odot $) this region lies within the
sensitivity band of space-based gravitational wave detectors such as eLISA. When interpreted
within the context of the super-massive black hole at the galactic center, Sgr A*,  this
implies that characteristic length scales of $41 \mu a s$ and $55 \mu a s$ and timescales
of $50 min$ and $79 min$ respectively should be associated with resonant effects if Sgr A* is non-spinning, while spin decreases these values by up to $\sim 32 \%$ and $\sim 28\%$.
These length-scales are potentially resolvable with radio VLBI measurements using the Event Horizon
Telescope.
We find that all low-order resonances are localized to the strong field region. In
particular, for distances $r > 50 R_s$ from the black hole, the order of the
resonances is sufficiently large that resonant effects of generic perturbations are not expected to lead to
drastic changes in the dynamics.
This fact guarantees the validity of using approximations based on averaging to model
the orbital trajectory and frequency evolution of a test object in this region.
Observing orbital motion in the intermediate region $50 R_s <r <1000 R_s$ is thus a
``sweet spot'' for systematically extracting the multipole moments of the central object
by observing the orbit of a pulsar -- since the object is close enough to be sensitive to
the quadruple moment of the central object but far enough away not to be subjected  to
resonant effects.

\end{abstract}
\maketitle

\section{Introduction}
\label{intro}
Super-massive black holes such as Sgr A*  at the center of our galaxy are at zeroth order mathematically idealized as Kerr black holes. In practice  this description is not complete due to a plethora of small perturbing effects which slightly alter the spacetime geometry.  In general these perturbations are small and well accounted for with canonical perturbation theory.  In the special case that the perturbation excites one of the intrinsic resonant structures of the spacetime's orbits, the effect may be larger than normally expected due to an anomalous transfer of energy and angular momentum that occurs during such a  perturbation.
Resonance phenomena are ubiquitous in any multi-frequency system. In celestial mechanics they strongly influence satellite dynamics and ring formation. Examples include  the gaps in the asteroid belt between Mars and Jupiter \cite{asteroid1} and the gaps in the rings of Saturn \cite{Saturn,Asteroid, rings}. Resonances are further intimately connected with dynamical chaos \cite{2001Natur.410..773M}.

As radio telescopes increase in sensitivity and collecting area we will be able to
resolve length-scales typical of resonant phenomena in the spacetime of the black hole at the center of our
galaxy. The Event Horizon Telescope is one such observational tool currently under 
development \cite{eventhorizontel}. Space based gravitational wave detectors such as eLISA
will be sensitive to in-spiral frequencies traverse the resonant regime and may observe 
shifts in the phasing of the gravitational waves emitted during the in-spiral of a compact object as it passes through the various resonant bands. X-ray, optical and infrared telescopes do not have the 
resolving power to image Sgr A* directly, but can potentially record flux variations 
from this region that may display timescales characteristic of resonant events.

This paper investigates resonant orbits in the Kerr metric expanding on the discussion in \cite{ourPRL}. The aim is to provide a complete characterization of the parameter space where resonant orbits occur as a function of
black hole spin and the orbital parameters. 
Since geodesic orbital motion in Kerr is completely integrable, it is akin to geodesic flow on a two-dimensional torus in phase space. Generic orbits are ergodic and sample the entire surface of the torus after a sufficiently long time. Low order resonant orbits however only trace out a simple, co-dimension one, curve on the torus. Some of the features of resonant orbital trajectories are illustrated  in~\cite{2012PhRvD..85b3012G,2014PhRvD..89h4036R,janna1, janna2, 2011MNRAS.414.3212H}.
By the  Kolmogorov-Arnold-Moser (KAM) theorem  which is discussed in Sec.~\ref{sec:KAM},  low order 
 resonant orbits are most likely to exhibit the non smooth anomalous behavior associated with a rapid change in the constants of motion  and the breaking of the resonant torus. Test particles entering a low order resonance often display subsequent dynamics with a sensitive dependence on initial conditions. 

To date a number of authors have studied resonant effects in Kerr-like metrics in the context of various forms of perturbations.  The effect of perturbations originating from adding a quadruple moment to the Kerr metric
has been quantified by exploring orbital motion in the Manko-Novikov metric \cite{Contopoulos2011, Contopoulos2012, Gair, JdB1, MGeyerThesis}. Perturbations from the presence of a disk were considered in Ref. \cite{Semerák01102012}, and the effects of the small mass' spin in \cite{spin1,spin2,spin3}. The features of traversing a resonance during an in-spiral, where the perturbation arises from the small mass' gravitational self-force, have been explored by \cite{prl, uchopol}, and the possibility of sustained resonance has been considered in \cite{PhysRevD.89.084033}.
Resonances involving one of the fundamental frequencies of the motion on the torus and the orbit's rotational frequency were studied in the context of enhanced gravitational recoil \cite{PhysRevD.83.104024,PhysRevD.90.044027}, and isofrequency orbits were discussed in \cite{isofrequ}.

Most of these studies have focused on a particular orbital trajectory or a small subset of parameters in a specific perturbed setting. The idea of this paper is to refrain from specializing to a particular perturbation and instead provide insights that apply to all types of resonant behavior. We will use tools such as the results of the KAM theorem that hold true regardless of the source of perturbation.
The results obtained here are thus robust in the sense that the time and length-scales of resonance effects for astrophysical applications are to be associated with properties of the underlying Kerr metric and resonance location rather than the details of the effect causing the perturbation.
The aim of this paper is to make the typical resonance time and length-scales accessible to the larger astrophysics community by means of easily evaluated formulas and tabulated results.

To explore the resonance effects  we describe the orbits in the Kerr metric using a set of variables adapted to the orbital geometry \cite{Schmidt} that reduce to the Keplerian orbital parameters in the Newtonian limit rather than the constants of motion associated with the spacetime's Killing fields. The properties of the Keplerian constants will be reviewed in Sec.~\ref{KeplerianCoords}. Plotting the location of resonances in terms of these variables immediately allows us to interpret the result as a physical location in the actual spacetime.

A resonant orbit occurs if the ratio of the characteristic radial, $\omega_r$, and longitudinal, $\omega_\theta$, frequencies is a rational fraction, \mbox{$\omega_r/\omega_\theta=n/m$} where \mbox{$ n, \ m \in \mathbb{N}$}. For a more extended discussion of the orbital geometry and frequencies see Sec.~\ref{KeplerianCoords}.  Most of the technical aspects of this paper deal with how to efficiently examine this expression and extract the physics. Closed form analytic expressions for the frequencies in terms of elliptic integrals have been presented by \cite{Schmidt,FujitaHikida} which serve as companions to this work.
Here, however, we opt in Sec. \ref{Sec:resonanceCond} to take advantage of a more symmetric representation of the elliptic functions appearing in the resonance condition and write them in terms of Carlson's integrals  \cite{NIST:DLMF, Olver:2010:NHMF, 1995NuAlg..10...13C}.  This allows us to identify the important parameters in the problem and exploit the identities associated with Carlson's integrals to manipulate the expressions.  In Sec.~\ref{SolSec} we consider solutions to the resonance condition. We first specialize to the weak field limit where we introduce the key properties of a ``resonant surface'' in the parameter space. We then give a number of exact analytic solutions to the resonance condition that can be used to describe resonances in the strong field region near the black hole. Finally, several low order (small $n+m$ value) resonant surfaces such as the $1/2$, $2/3$ and $3/4$ are evaluated numerically and compared to the analytic results and approximate formulae.

The breakdown of integrability around a resonance in ``almost''-Kerr spacetimes is often quantified by numerically generating Poincar\'{e} maps for a fixed energy $E$ and angular momentum component $L_z$. Associated with each Poincar\'{e} plot is a rotation curve which characterizes the frequency ratio as a function of  initial condition given a fixed $E$ and $L_z$.
In Sec.~\ref{Sec:ROT} we give a representative example of orbital breakdown around the $2/3$ resonance and analytically compute Kerr's rotation curve. We further provide expressions for finding the $E$ and $L_z$ values associated with a particular resonance.

The exact nature of a perturbed system's response in the region of
a resonance depends on the source of perturbation.
In Sec.\ref{Breakdown} we heuristically discuss how one would estimate the size of a perturbation required to see a dramatic change in dynamics. It is important to note that the KAM theorem does not guarantee the breakdown of integrability at any particular resonance. It merely states that if integrability breaks down it will occur first at the location of a low order resonant orbit \footnote{
Integrability could also break down 
 at a homoclinic orbit, e.g. the last stable orbit discussed in the Appendix~\ref{ISOapp}. However, this is of less observational interest than the resonances because it marks the transition to the plunge, where the nature of the motion changes drastically.}. Since this is true of all possible sources of perturbation, the cumulative effect of many sources of perturbation could result in a Saturn ringlike structure (see Fig.~\ref{SaturnsRings}) being established around the black hole. This and other potentially observable effects due to resonances are discussed in Sec.~\ref{astrophysics}.
We focus in particular on the galactic center, Sgr A*, as a possible observational realization of an extreme mass ratio in-spiral (EMRI). We note which detectors will be sensitive to chaotic orbits as well as the implications of regions were we can guarantee the absence of low-order resonances and in which we expect orbits to be approximately integrable. Regions that only contain high order resonances we consider to be the ``sweet spot'' for observationally determining the higher order multipole moments of the super massive black hole in the Milky way.

\section{KAM Theorem and importance of resonant orbits}
\label{sec:KAM}
\begin{figure}
\centering
\includegraphics[width = \columnwidth]{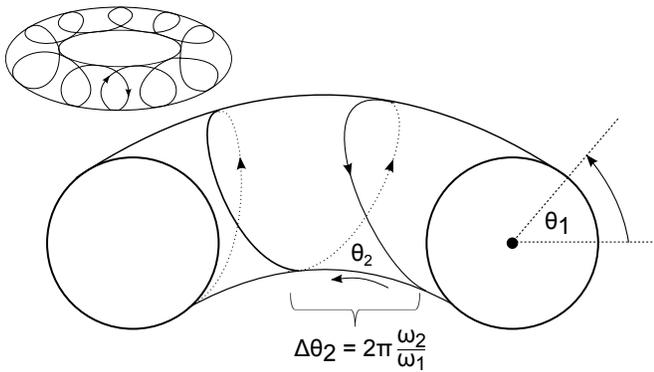}
\caption{
 The orbits in an integrable system with two degrees of freedom can be visualized as trajectories wrapping around a two dimensional torus in phase space with characteristic frequencies $\omega_1$ and $\omega_2$, relating to the angular advances in $\theta_1$ and $\theta_2$. For rational values of $\omega_2/\omega_1 = m/n$ the orbital trajectory will trace out a distinct path, wrapping $n$ times around the $\theta_1$ axis and $m$ times about the $\theta_2$ axis. For irrational values of $\omega_2/\omega_1$ on the other hand a trajectory will fill the surface of the torus densely.
\label{figtorus}}
\end{figure}
Bound geodesic motion in the Kerr spacetimes is integrable~\cite{Carter1968} since the
Hamiltonian $\mathcal{H}_K = 1/2 g_K^{\mu\nu} p_\mu p_\nu$, where $g_K^{\mu\nu}$ is the inverse Kerr metric and $p_\mu$ the test particle's four momentum, admits a full set of
isolating integrals. Two of these integrals result from the absence of explicit time and
azimuthal dependence in the Kerr metric functions, the third is due to the conservation
of rest mass and the fourth integral is known as the Carter constant~\cite{Carter1968}.
Integrability implies that action-angle variables can be defined. The phase space
is foliated by invariant level surfaces of the actions with the compact dimensions of
these surfaces diffeomorphic to a torus. Geodesic motion in an integrable system is thus
akin to geodesic flow on a torus.

To illustrate this idea
for the Kerr metric,  consider the reduced Hamiltonian, which is constructed by
replacing the conjugate momenta associated with the time and azimuthal symmetries
by their constant values
to obtain an integrable two degree of freedom  system with an effective potential \cite{JdB1}.  The main features of geodesic flow for such a  system are sketched in Fig.~\ref{figtorus}.
The trajectory on the torus is described by two characteristic frequencies, associated with the angles $\theta_1$ and $\theta_2$,  labeled $\omega_1$ and $\omega_2$, which in the Kerr metric correspond to the radial and longitudinal motions.
For rational values of $\omega_2/\omega_1$ the orbit will sample only a finite region of the torus before retracing its own path, while for irrational values of $\omega_2/\omega_1$ a trajectory will fill the torus densely. Orbits with rational frequency ratios involving large integers are very similar to irrational ones, it is only those with small integer ratios that are substantially distinct from the ergodic case.

When describing the astrophysical environment around a black hole such as Sgr A* we need to take into account a number of corrections to the mathematically idealized vacuum Kerr metric. In this case we are interested in the Hamiltonian
\be
\mathcal H = \mathcal H_K + \epsilon \mathcal H_1,\label{pertH}
\ee
where $\mathcal H_1$ is the perturbing Hamiltonian and $\epsilon$ is a dimensionless parameter characterizing the strength of the perturbation.
$\mathcal {H}_1 $  contains information about a possible accretion disk \cite{2008PhRvD..77j4027B,2012ApJ...759..130P,2011PhRvD..84b4032K}, other sources of matter \cite{2011PhRvD..83d4030Y} or dark matter \cite{PhysRevLett.83.1719, PhysRevD.88.063522},
structural deviations of the central black hole away from the Kerr metric (i.e. Bumpy
Black hole effects \cite{2004PhRvD..69l4022C, 2010PhRvD..81b4030V,2011PhRvD..83l4015J,2007PhRvD..75d2003B,2006CQGra..23.4167G}), the influence of modified gravity \cite{2011PhRvD..84f4016G,2011PhRvD..84f2003C}, or effects of the multipoles of
the small mass \cite{2009CQGra..26u3001B, spin1,2012PhRvD..86d4033S}. The exact nature of the perturbation does not concern us here. In what
follows we simply assume these modifications to be small and represent this by
considering the case where $\epsilon \ll 1$.

To quantify the effect of an arbitrary perturbation on the orbital motion and to find the regions where the impact of the perturbation will be greatest, we make use of the Kolmogorov--Arnold--Moser (KAM) theorem \cite{Arnold1963a, Moser1973}. The KAM theorem investigates the stability of \textit{near}-integrable systems and 
suggests that a torus associated with a rational ratio of frequencies will be destroyed
in the presence of perturbations. However, provided that the perturbation is small
enough, tori for which the ratio of associated characteristic frequencies are
\textit{sufficiently irrational} will remain stable and persist, although slightly
deformed, in the perturbed Hamiltonian  \cite{DetermChaos,Lichtenberg}. More specifically, consider the vector of frequencies  $ \bfomega $ in the unperturbed Hamiltonian and a vector of integers ${\bf k}$, and let $d$ denote the dimension of these vectors. The condition for resonance is ${\bf \bfomega \cdot k } =0$, which can generally be satisfied to arbitrary accuracy by choosing large integers for $\bfk$. When sufficiently large integers are necessary to satisfy the resonance condition the tori will be preserved, where the definition of \textit{sufficiently} is such that Arnold's criterion holds~\cite{Arnold1963a, Moser1973}
\begin{align}
|{\bf \omega \cdot k } | > K(\epsilon) \left (\sum_{i=1}^{d} |k_i|  \right)^{-(d+1)} .\label{eqKAM}
\end{align}
We will henceforth call $O_{\bf k}=\sum_{i=1}^d{|k_i|}$ the \textit{order} of the resonance. The factor $K(\epsilon)$ in Eq.
(\ref{eqKAM}) approaches zero as the perturbation vanishes, i.e.
$\lim_{\epsilon \to 0} K(\epsilon)\to 0$, but its functional form depends
on the nature of the perturbation. In a non-integrable Hamiltonian system, when $\epsilon\lesssim 1$, Eq.\eqref{eqKAM} suggests a hierarchy of resonant orbits of increasing order whose stability cannot be guaranteed.
These are the low-order resonances $1$, $1/2$, $1/3$, $2/3$, $1/4$, $1/5$, $3/4$, $2/5$,
$1/6$. We expect these tori to be destroyed first if the Hamiltonian is perturbed,
however, from Eq.\eqref{eqKAM} we cannot guarantee their destruction either. Changing
the Kerr metric's spin parameter is an example of a Hamiltonian perturbation to an
integrable Hamiltonian for which none of the lower order resonant tori are broken.

The destruction of resonant tori corresponds to the physical idea that energy transfer takes place most rapidly if the frequency of the driving force coincides with multiples of the internal frequencies of the system. Similarly, even without a direct input of energy, if a system is deformed the modes that could potentially be altered most are those whose frequencies are rationally related to other modes and which thus have the greatest potential to exchange energy and interact among themselves.

The study of torus destruction is not the subject of this paper. We do however give a heuristic discussion on how to estimate the size of the perturbation required for the onset of strongly chaotic dynamics in Sec. \ref{Breakdown}. The detailed calculation will differ depending on the characteristics of the perturbation.
The main focus in the following sections is to identify the regions in parameter and physical space where resonant dynamics are likely to occur. If they do occur the KAM theorem limits the impact to low-order resonances.

\section{Geodesic motion in the Kerr Metric}
\subsection{Physically motivated constants of motion}
\label{KeplerianCoords}
\begin{figure}
\centering
\includegraphics[width = \columnwidth]{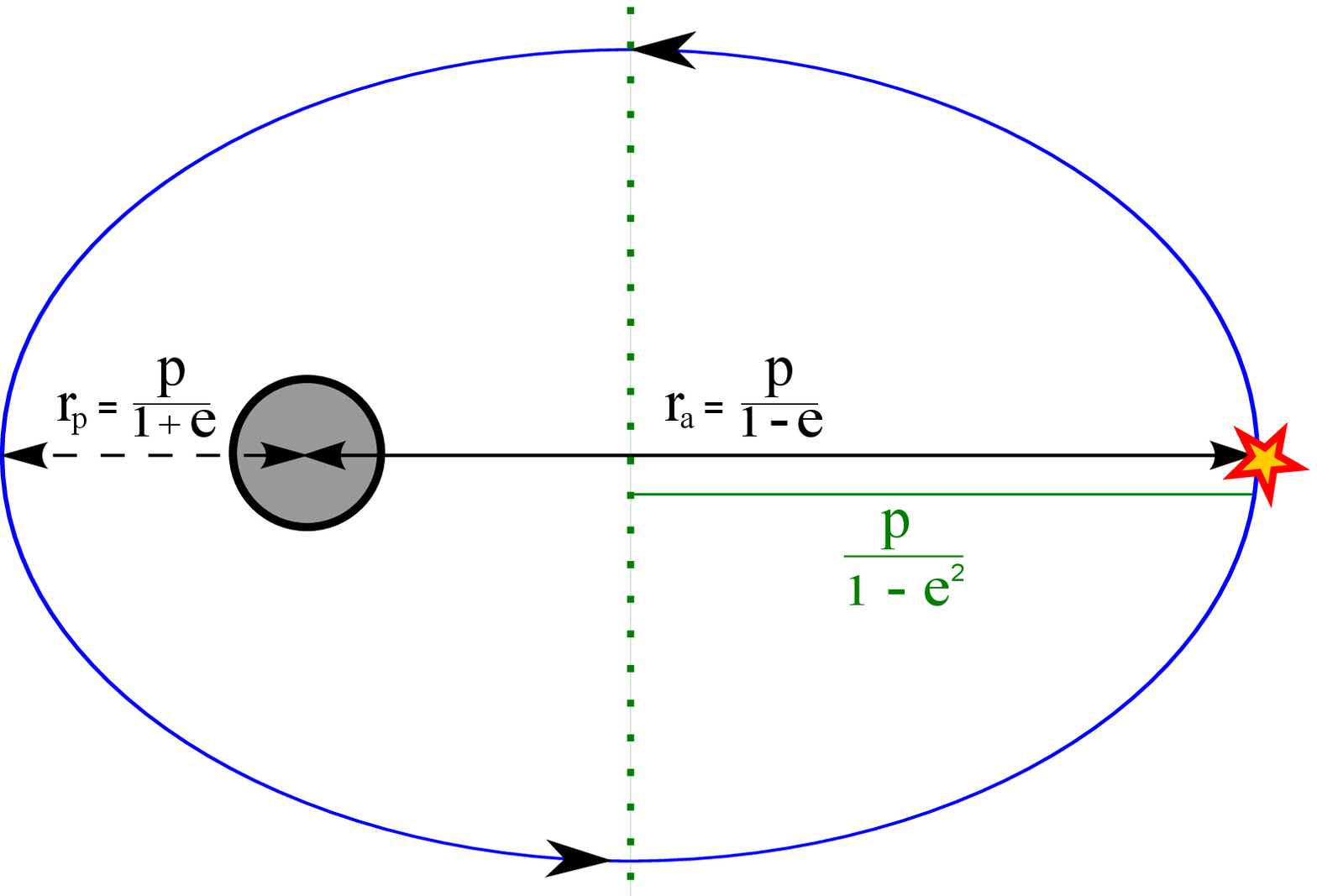}
\hspace{20pt}
\includegraphics[width = \columnwidth]{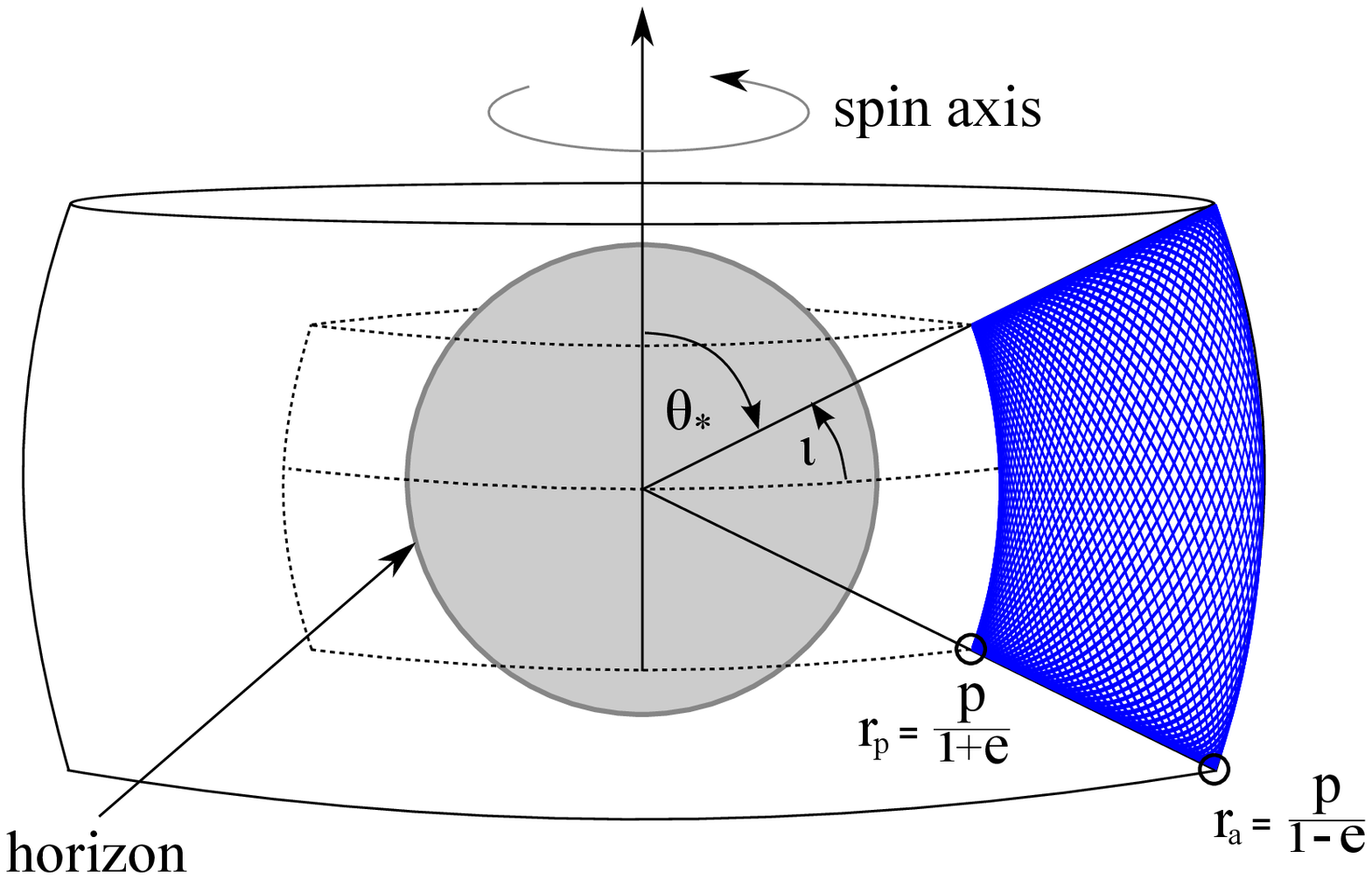}
\caption{\footnotesize{\textit{Top} Keplerian orbital parameters.
The eccentricity, $e$, is a measure of how elliptic the orbit it is.  
When \mbox{$e= 0$}, the orbit is circular  $e = 1$ the trajectory becomes parabolic. The semi-latus rectum, $p$, can be defined in terms of the eccentricity and the semi-major axis of the ellipse as is show in green in the figure. The point of closest approach $r_p$ is called the \textit{periastron}, while the most outlying point the orbit reaches is the \textit{apastron} denoted by $r_a$. \textit{Bottom} The orbital trajectory as shown in three dimensions. The third orbital parameter, namely  the maximum inclination angle \mbox{$ \iota = \pi/2-  \theta_{*}$}, is the angle with respect to the black hole's equatorial plane and $\theta_{*}$ is the minimum Boyer-Lindquist $\theta$ value attained.
}\label{figkep}}
\end{figure}

The  orbital motion of a bound trajectory of two bodies in Newtonian gravity is
described completely by an ellipse restricted to a plane. The manner in which this
ellipse is traversed is characterized by a single frequency, $\omega_\phi$. A schematic
representation of a typical elliptic orbit  and the  Keplerian variables used to
describe it is given in Fig. \ref{figkep}. By contrast in the Kerr metric bound orbits
are not restricted to a plane but are confined to a toroidal region whose shape is
characterized by the constants of motion, the energy $E$, the $z-$component of angular
momentum $L_z$ and Carter constant $Q$. For geodesics in Kerr, the rotational frequency
$\omega_\phi$ describing the rotational motion in the azimuthal direction is augmented by
two libration-type frequencies $\omega_r$ and $\omega_\theta$ which characterize motion in the radial and longitudinal
directions respectively. The bottom panel in Fig.~\ref{figkep} 
gives a schematic representation of the origin of the $\omega_r$ and $\omega_\theta$ 
frequencies associated with the orbit.

In the subsequent sections we will explore the location of the resonances for the
$\omega_r$ and $\omega_\theta$ frequencies in Boyer-Lindquist coordinates. Instead of
using the constants of motion $\{E/\mu,\ L_z/\mu, \ Q/\mu^2 \}$\footnote{The rest mass of the probe, $\mu$, is introduced here to ensure the constants of motion are dimensionless.} we will describe the
orbits using variables analogous to the Keplerian variables of classical celestial
mechanics, namely the eccentricity ($e$), sine of the maximum inclination angle (\mbox{$\sin \iota = \cos  \theta_{*}$}) and
semi-latus rectum ($p$) as illustrated in Fig.~\ref{figkep}.
 These are defined by writing the periastron or point of closest approach to the
 central object as \mbox{$r_p = p/(1+e)$}, the  apastron or furthest point the trajectory
 reaches as \mbox{$r_a = p/(1-e)$} and the turning point of the longitudinal motion as $\theta_*$.
The typical frequency of oscillations between  $r_a$ and   $r_p$ is described by $\omega_r$, whereas the longitudinal oscillations about the equatorial plane, 
$-(\pi/2-\theta_*)\leq \iota \leq\pi/2-\theta_*$, are described by $\omega_\theta$.

\subsection{Equations of motion}
\label{EOM}

For a test mass in orbit around a Kerr black-hole the equations governing the
radial and longitudinal motion, expressed in Boyer-Lindquist coordinates  $(t, r, \theta, \phi)$, are \cite{Carter1968}
\begin{align}
 \left( \f{dr}{d\lambda} \right)^2&= R(r),&
  \left(\f{dz}{d\lambda}\right)^2 =& \Theta(z),
\label{eom}
\end{align}
where $z=\cos(\theta)$ and we have chosen to parameterize the orbit in terms of a non-affine evolution parameter $\lambda=\int d\tau/(r^2+a^2 \cos^2\theta)$, rather than the proper time $\tau$, so that the radial and longitudinal equations decouple (this in fact just corresponds to working in the extended phase space). The radial and longitudinal potentials can respectively be expressed as
\begin{align}
R &= \left[(r^2+a^2)E-aL_z  \right]^2 - \Delta \left[\mu^2 r^2 + (L_z -a E)^2 +Q \right],
\label{Rpot}\\
\Theta &= Q(1-z^2) - \left[(\mu^2-E^2)a^2(1-z^2)+ L_z^2  \right] z^2.
\label{rthpot}
\end{align}
where 
$\Delta = r^2-2Mr+a^2$ and $a = S/M$ is the spin per unit mass. (We will henceforth use units where $M=1$.) The $R$ and $\Theta$ potentials  are quartic polynomials of their respective arguments and can equivalently be expressed in factored form as
\begin{align}
R&=-\frac{\beta^2}{a^2}(r-r_1)(r-r_2)(r-r_3)(r-r_4) \label{Rpotfac}\\
\Theta&= \beta^2 (z^2 - z_-^2)(z^2-z_+^2)          \label{rthpotfac}
\end{align}
where $\beta^2 = (\mu^2-E^2) a^2$. In Eq. \eqref{Rpotfac} we label the roots so that $r_1\geq r_2 \geq r_3 > r_4$ and in Eq. \eqref{rthpotfac} so that  $z_+\geq z_-$.
For bound orbits, Eq. \eqref{eom} dictates that the  $r(\lambda)$ and $z(\lambda)$ functions describing the orbital motion oscillate between  two of the roots of Eqs. \eqref{Rpotfac} and \eqref{rthpotfac} respectively. The generalized Keplerian variables are defined in terms of the roots of the potential functions as:
\begin{align}
r_1 &= \frac{p}{1-e}, & r_2 &= \frac{p}{1+e}, & z_-=\cos(\theta_*) 
\label{pet}\end{align}
When quantifying the resonance behavior in the subsequent sections
we would like to express the results entirely in terms of the variables $\{p,\ e,\ z_-\}$ rather than using $\{E,\ L_z,\ Q\}$.  The fact that the roots $r_3,\ r_4,$ and $z_+^2$ cannot  be viewed as independent functions but rather must be interpreted as  functions of the set of independent variables   $\{p,\ e,\ z_-\}$  complicates the calculation.
By comparing Eqs. \eqref{rthpot} and \eqref{rthpotfac} we can find
$z_\pm^2$ is given explicitly in terms of   $\{E, \ L_z, \ Q\}$  as follows
\begin{align}
\label{zpm2def}
z_{\pm}^2
&=
\frac{\left[(L_z^2+Q + \beta^2)
\pm \sqrt{(L_z^2+Q+\beta^2)^2 -4 Q\beta^2 }\right]}
{2\beta^2}.
\end{align}

Equating the coefficients of
$r$ in the two expressions for the radial equation, Eqs. \eqref{Rpot} and \eqref{Rpotfac} allows us to obtain the following expressions relating the constants
$\{E, \ L_z, \ Q\}$ to the roots of the factorization
\begin{align}
\frac{E^2}{\mu^2}
 &=1-\frac{2(1-e^2)}
{2p+\left(1-e^2\right) \varpi_+} , \notag\\
\frac{L_z^2}{\mu^2}&= \frac{2 p \left(p+2
   \varpi_+\right) -2 a^2
   \left(1-e^2\right) }{ 2 p+\left(1-e^2\right)
   \varpi_+}\notag\\
&+ \frac{2\left(  a^2\left(1-e^2\right)  -p^2\right) \varpi_{\times}}{a^2 \left(2 p+\left(1-e^2\right)
   \varpi_+\right)},
\notag\\
\frac{Q}{\mu^2}&= \frac{2  p^2  \varpi_{\times}}{a^2 \left(2 p+\left(1-e^2\right)
    \varpi_{+}\right)}
\label{condQEL}
\end{align}
where  we have
set $\varpi_+=r_3+r_4$ and $\varpi_{\times}=r_3 r_4$.
In addition the condition
\begin{align}
\frac{2 a E L_z}{\mu^2}&= a^2 +  \frac{ 2 \left(
 \varpi_{\times}
-2a^2 \right) (1-e^2)
 -2 p  \varpi_{\times}
 }{  \left(2
   p+\left(1-e^2\right) \varpi_{+}\right) }
\notag\\&
  \frac{-p^2
   \left( \varpi_{+}-2\right)+4 p  \varpi_+}{  \left(2
   p+\left(1-e^2\right) \varpi_{+}\right) }
\label{Condelz}
\end{align}
must also hold.
Squaring Eq. \eqref{Condelz} and then substituting in the expressions for $E^2$ and $L_z^2$, Eq. \eqref{condQEL} results in a quadratic equation for $\varpi_{+}$ and $\varpi_{-}$ in terms of  $p$ and $e$:
\begin{align}
 &\varpi_+ \varpi_{\times} \left[p (1-e^2) (p+4-a^2)+p^2(p-4 )-a^2 \left(1-e^2\right)^2
   \right]
\notag\\
&+ \varpi_{\times}^2 \left[p+e^2-1\right]^2
+2 p \varpi_{\times}\left[ \left(p-a^2\right) \left(p+e^2-1\right) \right]+\notag\\
&\frac{1}{4}\varpi_+^2 \left[a^4 \left(1-e^2\right)^2-2 a^2 \left(1-e^2\right) p
   (p+4)+(p-4)^2 p^2\right]\notag\\
&+ p \varpi_+ \left[a^2 \left(1-e^2\right)(a^2+ p)
   -a^2 p
   \left(p+4\right)-(p-4) p^2\right]\notag\\
&+ p^2
   \left(a^2-p\right)^2 =0. \label{wptquad}
\end{align}
If $z_-^2 \neq 0$  we can use
Eqs. \eqref{zpm2def} and \eqref{condQEL}
to rewrite $z_+^2$ as
\begin{align}
z_+^2 = \frac{ p^2 \varpi_{\times}}{a^4(1-e^2) z_-^2}.
\label{zp2}
\end{align}
We will always treat the $z_-^2=0$ or $Q=0$  limit of orbits restricted to the equatorial plane separately.
Using Eqs.  \eqref{zpm2def}, \eqref{condQEL}  and \eqref{zp2}  we can further show that
\begin{align}2 a^2 p \varpi_+ z_-^2=\left(a^2 \left(1-e^2\right) z_-^2-p^2\right) \left(a^2 z_-^2-\varpi_{\times}\right)
.
\label{wpcond}
\end{align}
Eq. \eqref{wpcond} is a linear condition in  $\varpi_{+}$ and $\varpi_{\times}$
which, in conjunction with \eqref{wptquad},
implicitly determines $\varpi_{+}(p, e, z_{-}) $ and $\varpi_{\times}(p, e, z_{-}^2) $  and thus the roots $r_3$  and $r_4$ in terms of $\{p, e, z_{-}\}$.
Substituting  Eq. \eqref{wpcond} into
 \eqref{wptquad}  eliminates $\varpi_+$ and yields a  quadratic equation in $\varpi_{\times}$.
As a result a closed form expression can easily be found for $\varpi_{\times}$ and subsequently $\varpi_{+}$. We will not give the expressions here and continue to work with the implicit quantities  $\varpi_{+}$ and $\varpi_{\times}$, substituting their actual values only at the end of the calculations. The two solutions that result from the quadratic equation can be interpreted as test masses that either co-rotate or counter-rotate with respect to the spin, $a$, of the black-hole.
 Orbits that co-rotate with the black-hole  ($L_z$ has the same sign as $a$) are called \textit{prograde}  and those that counter-rotate ($L_z$ has the opposite sign to $a$) are \textit{retrograde} orbits.
For a given $\{ p, e, z_-\}$ the
prograde orbit's angular momentum is higher than that of the retrograde orbit. On the other hand prograde orbits have lower orbital energy than their retrograde counterparts  \cite{Schmidt}.

One special set of orbital parameters is the case when the roots satisfy $r_3=r_2$, which corresponds to the innermost stable orbit (ISO) separating stable bound orbits from those that plunge into the black-hole. For a given eccentricity and longitudinal parameter $z_-$  the semi-major axis satisfying this condition demarcates the smallest value of $p$ at which a stable bound orbit can exist. We shall explicitly solve for the ISO  for all values of $a$, $e$ and $z_-$ in Appendix~\ref{ISOapp} and use it as a comparative benchmark for the location of resonant orbits in the subsequent sections.

\section{The resonance condition}
\label{Sec:resonanceCond}
In this section we begin to characterize the orbits which will exhibit resonant behavior.
We are interested in the parameter values for which $\omega_r$ and $\omega_\theta$ are commensurate. Given relatively prime integers $m$ and $n$ we seek the surface in the three dimensional parameter space spanned by $p$, $e$ and $z_-$ where
\begin{align}
m\omega_r = n\omega_{\theta}.
\end{align}
This is equivalent to saying that the time it takes the longitudinal motion to traverse exactly $n$ times between its turning points is equal to the time it takes the  radial motion to traverse $m$ times between its turning points. For Kerr geodesics $m\geq n$ since the radial frequency is always the smallest of the three frequencies. This translates into the following integral condition
\begin{align}
m \int_{-z_-}^{z_-} \frac{  dz }{\sqrt{\Theta}} = n \int_{r_2}^{r_1} \frac{dr}{\sqrt{R}}\cM{.}
\end{align}
Substituting Eqs. \eqref{Rpotfac} and \eqref{rthpotfac}
we find that this is equivalent to the condition
\begin{align}
a n \int_{r_1}^{r_2} \frac{dr}{\sqrt{ (r_1-r)(r-r_2)(r-r_3)(r-r_4) } }= \notag\\
-m \int_{-z_-}^{z_-} \frac{dz}{\sqrt{ (z^2 - z_-^2)(z^2-z_+^2)  }  } \cM{.} \label{RESOCON1}
\end{align}
The subject of the rest of the paper is to characterize the solutions to this equation.
The strategy is to express both the radial and longitudinal integrals in their most symmetric form using Carlson's integrals  \cite{NIST:DLMF, Olver:2010:NHMF, 1995NuAlg..10...13C} . Carlson's integral of the first kind is defined to be
\be
R_F(\alpha,\beta,\gamma) = \frac{1}{2} \int_0^\infty \frac{dt}{\sqrt{(t+\alpha)(t+\beta)(t+\gamma)}}.
\ee
In Appendix \ref{app:Carlson} we list a number of identities and rapidly converging approximation techniques that make Carlson's integrals a valuable analytic tool for characterizing the resonant surfaces.
Using Eq. \eqref{CI1} of  Appendix \ref{app:Carlson} we can rewrite Eq. \eqref{RESOCON1} as
\begin{align}
a n R_F(0, ( r_2  -r_3) (r_1  -r_4),        ( r_2  -r_4)(r_1  -r_3) ) =  \notag\\
-m R_F( 0,  (z_-+z_+)^2,     ( z_- -z_+)^2) .\label{RESOCON2}
\end{align}
 This expression can be further simplified using the identity \eqref{Rfxysq} to rewrite the right hand side  and the fact that the equations are homogeneous  \eqref{HomoGen} to absorb the constant factor.  We shall refer to the resulting equation,
\begin{align}
 R_F(0, ( r_2  -r_3) (r_1  -r_4),        ( r_2  -r_4)(r_1  -r_3) ) =  \notag\\
 R_F( 0,  \kappa a^2 (  z_+^2-z_-^2 ),   \kappa a^2     z_+^2),
\label{RESCONDCARL}
\end{align}
as {\it the resonance condition} and explore its properties by studying  various limiting cases. In this expression we have defined the parameter $0<\kappa<1$ to indicate  which resonance we are considering,
\begin{align}
\kappa = \frac{n^2}{m^2}.
\end{align}
In the subsequent section we will explore all the qualitative features of a resonance by examining an easily evaluated approximation to Eq. \eqref{RESCONDCARL} for large $p$. We then give a number of formulae valid in the region near the blackhole in special cases.

\section{Solutions to the Resonance Condition}
\label{SolSec}
When seeking  solutions of Eq.~\eqref{RESCONDCARL} it is convenient to rewrite it in terms of a rapidly converging series.  This series allows us to identify  the three important parameters in the problem. The first sets the overall scale and a rough location of the resonance. The remaining two are expansion parameters $< 1$ that determine the more subtle structure of the resonance surface. We give explicit expressions for theses parameters in terms of the variables introduced in Sec. \ref{EOM}. Next, we evaluate the series in the large $p$ limit to obtain a simple analytic model which illustrates the important features of any resonance. We then turn to the astrophysically more interesting strong field region where the low-order resonances occur and give a number of exact analytic formulas for special cases. We conclude the section by
  numerically computing the detailed behavior of the $2/3$ resonance and compare our analytic results and approximations to the numerical solutions.

\subsection{General series expansion }
\label{seriesExp}
The resonance condition
\eqref{RESCONDCARL}
can be rewritten in the form
\begin{align}
R_F(0,y_1+\delta_1 , y_1-\delta_1)=R_F(0,\kappa(y_2+\delta_2) , \kappa(y_2-\delta_2)). \label{RfSpre}
\end{align}
where
\begin{align}
y_1&= \frac{p^2-p(r_3+r_4)   }{1-e^2} + r_3 r_4, &\delta_1&= \frac{e p (r_4-r_3)}{1-e^2}, \notag\\
y_2 &=\frac{a^2 }{2} (2z_+^2-z_-^2), &  \delta_2 &= -\frac{a^2 z_-^2}{2}.
\end{align}
It will be shown below Eq. \eqref{paraExpand} that $\delta_i/y_i\ll 1$ for all physically interesting parameters. As a result, each side of Eq. \eqref{RfSpre}   can be expanded in $\delta_i\ll y_i$, using the rapidly converging seriesof Eq.  \eqref{taylor}. 
Squaring the resulting expansions, moving all the
 terms containing the small parameters $\delta_i/y_i$ to one side, and
re-expanding the result, we obtain the equation
\begin{align}\frac{y_1}{\kappa y_2} = & 1 + \frac{3}{8} \left(\frac{\delta_1^2}{y_1^2}- \frac{\delta_2^2}{y_2^2} \right) -\frac{9}{64} \frac{\delta_1^2 \delta_2^2}{y_1^2y_2^2}
 \notag\\
&+\frac{123}{512}  \left(\frac{\delta_1^4}{y_1^4}-  \frac{\delta_1^4}{y_1^4} \right)
+ O( \frac{\delta^6}{y^6} ).
\label{expansion}
\end{align}

In terms of the variables introduced in Sec. \ref{EOM},
the three quantities that enter the expansion of the resonance condition \eqref{expansion}
are,
\begin{align}
\frac{y_1}{y_2} &=  \frac{2 a^2 z_-^2 \left((1-e^2 )\varpi_{\times}+p^2-p \varpi_+\right)}{ \left(a^4
   \left(e^2-1\right) z_-^4+2 p^2 \varpi_{\times}\right)}  \notag\cM{,}\\
\frac{\delta_1}{y_1} &= \frac{-e p \sqrt{\varpi_+^2-4 \varpi_{\times}}}{\left(1-e^2\right) \varpi_{\times}+p^2-p \varpi_+} \notag\cM{,}\\
\frac{\delta_2}{y_2} &= \frac{\left(1-e^2\right)a^4 z_-^4}{(1-e^2)a^4 z_-^4-2 p^2
   \varpi_{\times}}\cM{.}
\label{paraExpand}
\end{align}
The first term $y_1/y_2$ ultimately sets the overall scale of $p$ at which a particular resonance occurs. Recall that the parameters $\{z_-,\ e, \ a,\ \kappa \} \in [0,1]$. By examining Eq.~\eqref{paraExpand} one can further verify that the $\delta_i/y_i$ terms are always less than unity ensuring the convergence of the series in Eq. \eqref{expansion}.
The parameters $\delta_1/y_1$ and $\delta_2/y_2$ vanish when $e=0$ and $z_- =0 $ respectively. The special limiting case when both these conditions hold, allows us to find an exact analytic result that is valid in all regions of the spacetime. We shall examine this special case in Sec.~\ref{eqcircle}.
However before we do so, it is instructive to examine the properties of resonances that occur at large $p$ values. The features we explore in this limit qualitatively capture the
 characteristics of resonances in general.

\subsection{Anatomy of a resonance in the weak field limit $p\rightarrow \infty$  }
\label{pinfLimit}

\begin{figure*}
\includegraphics[width = \textwidth]{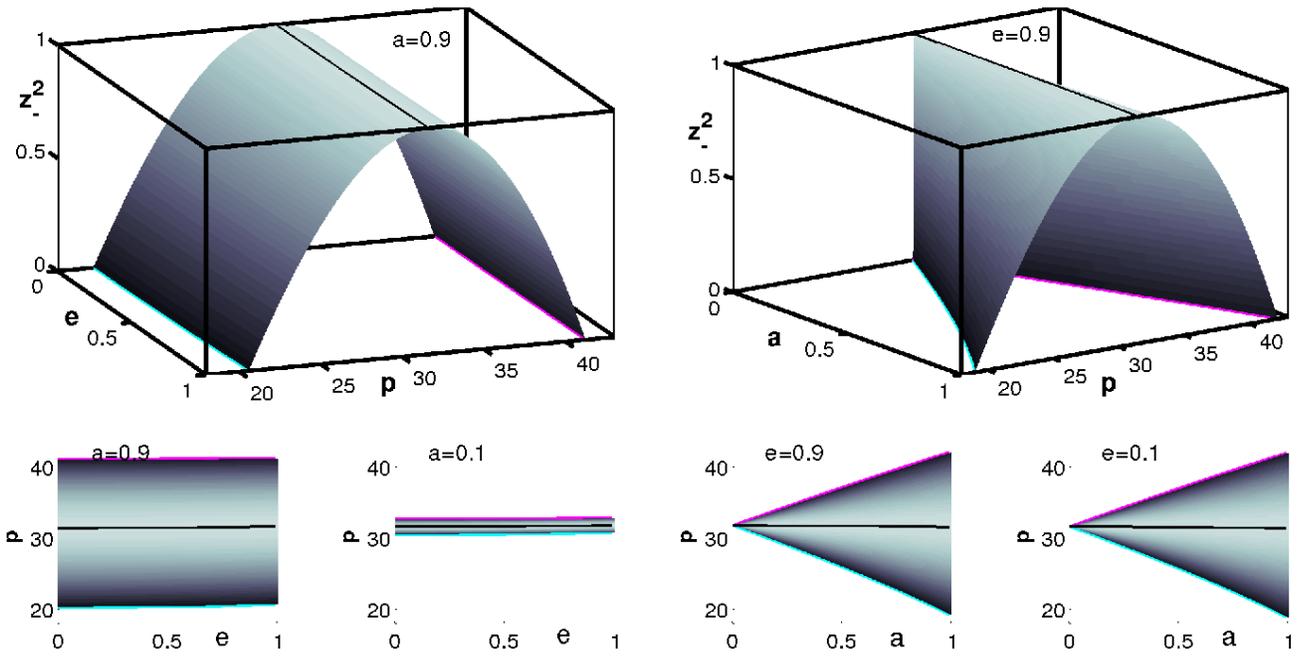}
\caption{\footnotesize{
Graphical representation of the approximate resonance condition \eqref{approxReso} or equivalently \eqref{z2meq} plotted for the  $\kappa=(9/10)^2$ resonance. 
The eccentricity dependence is shown in the three plots on the left. 
The top left plot displays the typical arch shape seen for all resonances. Here the arch is centered around $p^*\approx 31.6$ reaching a maximum value of $z_-=1$ at $p=p_{polar}$ given in Eq. \eqref{ppolar} and indicated by a dark line on the plot. The lines where  arch intersect the $z_-=0$ plane are  indicated in magenta (right) and cyan (left) and show the maximum (retrograde) and minimum (prograde) values $p$ attains for a fixed $e$. Note that the resonance shape for a fixed spin is very weakly dependent on the eccentricity of the orbits. 
The spin dependence of the approximate resonance condition is shown in the three plots on the right for  $e=1/10$ and $e=9/10$.
Observe that as $a\rightarrow 0$ the arch pinches off to a line at
 $p=p_{polar}$, Eq. \eqref{ppolar}. The maximum arch width occurs for a maximally spinning black hole  $a=1$ as predicted by Eq. \eqref{pretropro}.
\label{res0304eall}}}
\end{figure*}

In the weak field limit, when $p\rightarrow \infty$, the dominant terms in the expansion of $\varpi_{\times}$ and $\varpi_+$ found by solving Eqs.\eqref{wpcond} and \eqref{wptquad}
are,
\begin{align}
\varpi_{\times}&= a^2z_-^2\left(1+\frac{4}{p} \pm \frac{8a \sqrt{(1-z_1^2)  } }{p^{3/2}}  \right)
+O\left(\frac{1}{p^2}  \right) \notag \cM{,}\\
\varpi_{+}
&=2+\frac{8+2 a^2
   (1-z_-^2)}{p}
 \notag\\
&\pm 4 a \sqrt{\frac{ 1-z_-^2}{p}} \left(1+
   \frac{6 }{p}     \right) + O\left(\frac{1}{p^2}\right)\cM{.} \label{varpipinf}
\end{align}
Substituting Eq. \eqref{varpipinf} into Eq. \eqref{paraExpand} and simplifying the result gives the dominant
behavior of the three essential parameters in the resonance condition
\begin{align}
 \frac{y_1}{y_2} &=1-\frac{6}{p}  \mp
  \frac{12a \sqrt{1-z_-^2}  }{p^{3/2}} \notag\\
&-\frac{3 a^2 \left(\left(e^2-5\right) z_-^2+4\right)}{2 p^2}+O\left(\frac{1}{p^{5/2}}\right)
     ,
\label{rscale}
\end{align}
\begin{align}
\label{smallpara}
\frac{\delta_1}{y_1}&=-\frac{2 e \sqrt{1-a^2 z_-^2}}{p} \mp
 \frac{4 a e}{p^{3/2}}\sqrt{\frac{z_-^2-1}{a^2
   z_-^2-1}} +O\left(\frac{1}{p^{2}}\right) \notag\\
\frac{\delta_2}{y_2}&=
-\frac{a^2 \left(1-e^2\right)
   z_-^2}{2 p^2} +O\left(\frac{1}{p^{3}}\right) \cM{.}
\end{align}
The  small $\delta_1/y_1$  and $\delta_2/y_2$ parameters scale with $1/p$ and $1/p^2$ respectively indicating that they become almost negligible for large $p$ and reaffirming the choice of $\delta_i$ as a expansion parameter.

To illustrate the basic properties
of  the resonance condition
we substitute  Eqs. \eqref{rscale} and
\eqref{smallpara} into Eq. \eqref{expansion}
 and  keep terms up to $O(p^{-2})$.
The resulting approximate resonance condition,
\begin{align}
\pm 24 a
   \sqrt{p(1-z_-^2)}=2 p^2 (\kappa-1)+12( p+a^2) +3 e^2 \kappa    \notag\\
-3 a^2 z_-^2 \left(e^2 (\kappa-1)+5\right), \label{approxReso}
\end{align}
is valid for large $p$ values only. However, this weak field approximation demonstrates all the qualitative  properties of resonant surfaces and gives a good approximation even relatively close to the black-hole. The precise manner in which Eq.~\eqref{approxReso} breaks down for low order resonances is numerically explored in Sec.~\ref{Numerical2o3}.

To build our intuition of the typical features of resonant surfaces and their dependence on the parameters  $a, e, z_-, \kappa$  and $p$ we analyze Eq. \eqref{approxReso} in detail.
For quasi-circular orbits and vanishing black-hole spins $(a,e)\to 0$, the resonances  occur at
\begin{align}
p^*=p(a=0,e=0,z_-,\kappa) = \frac{6}{1-\kappa}.
\end{align}
For a given $\kappa$, this value of $p^*$ sets the general mean radius in physical space (measured in units of $GM/c^2$) about which all the interesting features of a resonance occur.
This is a robust result that remains an exact analytic solution even in the region near the horizon, as we will prove in Sec. \ref{eqcircle}. For a fixed integer $m$, resonances with $\kappa=\left[(m-1)/m\right]^2$ correspond to the maximum resonance radius given by $p=6m^2/(2m-1)$.
Resonances  with $n<m-1$ occur at a  radius less than that associated with $n=m-1$.   The maximum $p$ associated with a denominator $m$ thus scales linearly with $m$ for large values of $m$. 

In the limiting case $a\to 0$ the dependence on eccentricity is
\begin{align}
p(a=0,e,z_-,\kappa)= \frac{p^*}{2} \left(1+\sqrt{1+e^2\left( \frac{p^*-6}{p^{* 2}}\right)}\right).
\end{align}
Since $e\leq 1$ and $p^*>6$ we see that the effect of eccentricity on the resonance location is small. We also observe that in the case $a\to 0$ the location of the resonance becomes independent of $z_-$.

We now examine  the general spinning case.
Squaring both sides of Eq. \eqref{approxReso} results in a polynomial condition that is quartic in $p$ and quadratic in $z_-^2$.
We choose to analyze the solution surface by specifying the $z_-^2=z_-^2(a,e,p,\kappa)$ for a fixed $\kappa$ rather than explicitly working with the quartic roots associated with $p$.  The appropriate expression for $z_-^2$ is
\begin{align}
z_-^2 =\frac{4 a^2 + e^2  -\frac{6e^2+4p^2}{p^*}
  }{ a^2 \left(5-\frac{6
   e^2}{p^*}\right)}
-\frac{12 p \left(\frac{2
   e^2}{p^*}+1\right)}{ a^2 \left(5-\frac{6
   e^2}{p^*}\right)^2}
\notag\\
+\frac{8}{a^2}
\sqrt{\frac{ p \left[a^2 -e^2 +\frac{6 e^2(1-a^2)+4p^2}{p^*}    \right] }{    \left(5-\frac{6
   e^2}{p^*}\right)^3 }
- \frac{4 p^2 \left(1-\frac{6
   e^2}{p^*}\right)}
{  \left(5-\frac{6
  e^2}{p^*}\right)^4     }}
.\label{z2meq}
\end{align}
This function is depicted in 
Fig.~\ref{res0304eall} 
for a fixed spin parameter of $a=9/10$ and integer ratio $\kappa=(9/10)^2$.  It has the shape of a parabolic arch centered around $p^* \approx 31.6$.  
Furthermore, the qualitative features that will be discussed here are characteristic for all resonances.
The function given in Eq. \eqref{z2meq} has a maximum value of $z_-^2=1$
which occurs when $p_{polar}=p(a,e,z_-=1,\kappa)$ has the value
\begin{align}
\frac{p_{polar}}{p^*}&= \frac{1}{2}  \left(1+ \sqrt{1 +\frac{ \left(e^2-a^2\right)}{p^{*}}     -\frac{6 \left(1-a^2\right)
   e^2}{p^{*2}}}\right).
\label{ppolar}
\end{align}
Since $\{e,\ a\}  \in [0,1]$ and $p^*\gg 8$, the maximum only deviates by a few percent from the $p^*$ value as the spin and eccentricity deviate from zero.
The analytic value of the  maximum given by Eq. \eqref{ppolar} is plotted as a dark line in 
Fig.~\ref{res0304eall}.
When $z_-^2 <1$ there are two possible values of $p$ that lie on the resonance sheet for a given eccentricity:
the resonance for a retrograde  orbit $p_{-}>p_{polar}$ and the resonance for a prograde orbit which occurs closer to the black hole $p_{+}<p_{polar}$. The sign in the naming convention of retrograde and prograde orbits relates to the sign of the  product of the  angular momentum and the spin of the black hole ($a L_z$), and not the orbit's relative position with respect to $p_{polar}$.

As $z_-$ decreases and the resonance surface moves from the polar towards the equatorial region, the influence of spin becomes increasingly important and the distance  $p_{-}$ to $p_{+}$ monotonically increases. The expression for $p_-$ and $p_+$ can easily be found in closed form  by substituting $z_-=0$ into
 Eq. \eqref{approxReso} and solving the resulting quartic for $p$. However, since the results are messy and add little to the discussion we do not give the general results explicitly and merely  plot these curves in 
Fig.~\ref{res0304eall}. To benchmark the size of the arch we consider the limit of vanishing eccentricity and inclination and obtain 
\begin{align}
\frac{p_{\pm}(a,e=0,z_-=0,\kappa)}{p^*} =
\frac{1}{2} \left(1+\sqrt{1 \mp \frac{4
   a}{\sqrt{p^*}}}  \mp \frac{2 a}{\sqrt{p^*}}\right)
.\label{pretropro}
\end{align}
The maximum span of the arch occurs for a maximally spinning black hole, $a=1$. For lower spin values a good approximation of the span of the arch is \mbox{$(p_--p_+) \approx 4a\sqrt{p^*}(1+a^2/p^*) $}. The lowest order correction to Eq. \eqref{pretropro} with respect to eccentricity is $e^2(p^*-6)/(4p^{*2})$ and is the same for both pro and retrograde orbits.

Having thus explored the basic features of a resonance for a given spin parameter $a$ and observed the weak dependence of these features on eccentricity,
 we will now choose a representative eccentricity and then study the spin dependence. 
The right-hand three panels of Fig.~\ref{res0304eall}
show the $\kappa=(9/10)^2$ resonance surface for eccentricity values of $e=9/10$ and $1/10$ as a function of black hole spin and $p$.
As predicted by Eq. \eqref{pretropro} the arch-width exhibits a strong spin dependence. The arch's inverted $'U'$ profile pinches off to a single column $'I'$ profile at $p=p_{polar}$ (Eq. \eqref{ppolar}) when $a\rightarrow 0$. This indicates that resonances in the nonspinning limit become independent of inclination because the longitudinal frequency degenerates to the $\phi$-frequency in this case. As the black-hole's spin increases from zero the opening width of the arch between the pro- and retrograde branches increases until a maximum arch-width is attained at $a=1$. The result is  a $'V'$-shaped footprint of the arch in the $p-a$ plane, with the $'V'$ profile's vertex corresponding to $a=0$. The inclination dependence
of the resonance surfaces can simply be characterized as the monotonic closing off of the $'V'$  profile's pro-and retrograde branches with increasing inclination until they merge into  a single line forming the arch's spine at $p_{polar}$.

This completes our discussion of resonances in the weak field limit. The features described here and the $'U' -'V' - 'I'$ transitions are characteristic of all resonances. The actual values of the
resonant surface of the true resonance condition begin to deviate from our weak field model as the black hole is approached.  The largest deviation occurs in the equatorial limit, where the effect of spin is most marked. In the polar regions, the weak field resonance condition remains a remarkably accurate approximation to the true resonance surface.   In Sec.~\ref{Numerical2o3}  we numerically characterize several low order resonances and provide a quantitative comparison with the approximate results obtained in this section.

As we shall see next, in the strong field region it is
possible to obtain exact analytic results for the $'V'$ equatorial footprint for $e=0$.  Since the resonant surface depends very weakly on $e$ this result is a good indicator for all resonant behavior.

\subsection{Exact solutions to the resonance condition in special cases}
\label{eqcircle}
\begin{figure*}[t]
\includegraphics[width = \textwidth]{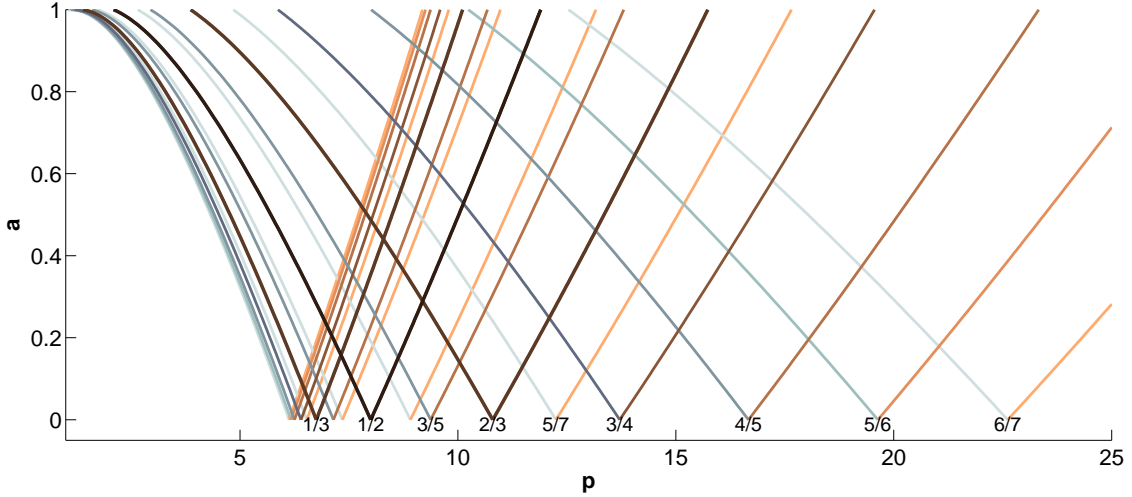}

\caption{The location of orbits with resonant frequencies in the limiting case of $e=0$ and $z_-=0$ as a function of spin, $a$ and semi-latus rectum, $p$ \cite{ourPRL}.
Resonances are labeled at their vertex by the rational ratio $n/m$. All resonances with $m\leq 7$ are shown. For $a=0$ the prograde and retrograde branches are degenerate at $p=p^*$, as the spin increases the retrograde branch leans right (copper tinge) and the prograde branch leans left (blue tinge). In general lower order resonances are colored more darkly than their higher order counterparts.
Note the accumulation of resonances as the strong field region is approached   \label{fig:spine0}}
\end{figure*}
\begin{table*}
\begin{tabular}{|c|c|c|c|ccc|} \hline
Resonance & Location  $\left[GM/c^2  \right]$& Spin splitting   &  Period $\left[GM/c^3\right]$& Galactic &center:& Sgr A* \\ \hline \hline
$\sqrt{\kappa}=n/m$& $p^*=6/(1-\kappa)$&Max$[(p_1-p_+)/p*]$& $T=2\pi p^{*3/2}$& $p*$  [$\mu a s$] & $T$ [$min$] & $f$ [$10^{-4}Hz$]  \\ \hline
 ISCO& 6 & 1.33 &92.3 &
 30.6  & 32.7 & 5.10 \\ \hline
\bf 1/2 & \bf 8  & \bf 1.22 & \bf 142.1 & \bf 40.9  &\bf  50.3
    &\bf 3.31  \\ \hline
\bf 1/3 & \bf 27/4 = 6.8 &\bf 1.29 & \bf 110.2 & \bf 34.5
 & \bf 39.0  & \bf 4.27  \\
 \bf 2/3 & \bf 54/5 = 10.8 & \bf 1.10 & \bf 223.0 & \bf 55.2  & \bf 78.9  & \bf 2.11  \\ \hline
 1/4 & 32/5 = 6.4 & 1.31 & 101.7 & 32.7
 & 36.0  & 4.63  \\
 3/4 & 96/7 = 13.7 & 1.00 & 319.1 & 70.1  & 112.9   & 1.48  \\ \hline
 1/5 & 25/4 = 6.3 & 1.32 & 98.2 & 31.9     & 34.7   & 4.80  \\
 2/5 & 50/7 = 7.1 & 1.27 & 119.9 & 36.5
    & 42.4   & 3.93  \\
 3/5 & 75/8 = 9.4 & 1.16 & 180.4 & 47.9  & 63.8   & 2.61  \\
 4/5 & 50/3 = 16.7 & 0.92 & 427.5 & 85.1 & 151.3   & 1.10  \\ \hline
 1/6 & 216/35 = 6.2 & 1.32 & 96.3 & 31.5 & 34.1   & 4.89  \\
 5/6 & 216/11 = 19.6 & 0.86 & 546.7 & 100.3    & 193.5   & 0.86  \\ \hline
 1/7 & 49/8 = 6.1 & 1.33 & 95.2 & 31.3     & 33.7   & 4.94  \\
 2/7 & 98/15 = 6.5 & 1.30 & 104.9 & 33.4 & 37.1   & 4.49  \\
 3/7 & 147/20 = 7.4 & 1.26 & 125.2 & 37.5  & 44.3   & 3.76  \\
 4/7 & 98/11 = 8.9 & 1.18 & 167.1 & 45.5    & 59.1   & 2.82  \\
 5/7 & 49/4 = 12.3 & 1.05 & 269.4 & 62.6  & 95.3   & 1.75  \\
 6/7 & 294/13 = 22.6 & 0.80 & 675.8 & 115.5
    & 239.1   & 0.70  \\ \hline
\hline
\end{tabular}
\caption{Time and length-scales associated with low-order resonances depicted in
  Fig.~\ref{fig:spine0} .  This table gives the values for the $e=0$, $a=0$, $z_-=0$  vertices seen in Fig.~\ref{fig:spine0},
 first in dimensionless units and subsequently in physical units for the special case of the Galactic center, Sgr A*.   Lower order resonances, shown in bold in this table, are most likely to have observationally detectable dynamics.
\label{TabRes}
}
\end{table*}

In this section we explore easily evaluated exact solutions
to the resonance condition of Eq. \eqref{RESCONDCARL} that can be used to characterize the resonant behavior near the black hole.
The case we will consider first is the limit of circular equatorial orbits, i.e.
$e\to 0$ and $z_-\to 0$. As remarked in Sec.~\ref{seriesExp} this case sets
the parameters $\delta_1/y_1=\delta_2/y_2=0$  in Eq. \eqref{RfSpre} and
thus a valid solution to the resonance condition is found when
\begin{align}
\frac{y_1}{y_2}=\kappa.
\label{Qe0recond}
\end{align}
Note that in evaluating this case we will not be resorting to
Eq. \eqref{paraExpand} that was derived using Eq. \eqref{zp2} which assumed that $z_-^2\neq 0$. Instead we return to Eq.~\eqref{zpm2def} and observe that $z_-=0$ if and only if $Q=0$. By Eq.~\eqref{condQEL} we see that $Q=0$ implies $\varpi_\times =0$. The simplified version of Eq.~\eqref{condQEL} is
\begin{align}
\frac{E^2}{\mu^2}
 &=1-\frac{2}
{2p+ \varpi_+} , &
\frac{L_z^2}{\mu^2}&=
4p- 2\frac{a^2+3p^2}{2p+\varpi_+}
\cM{.}\label{condQELQ0}
\end{align}
Substituting Eq.~\eqref{condQELQ0} into Eq.~\eqref{zpm2def}  gives an expression for
$z_+$,
\begin{align}
z_+^2=\frac{p(p+2\varpi_+)}{a^2}\cM{.}
\end{align}
Setting $\varpi_{\times}=0$ in Eq. \eqref{Condelz} results in a quadratic equation for $\varpi_+$ which has the following roots,
\begin{align}
\varpi_+ = \frac{2p(a\pm\sqrt{p})^2   }{p(p-4)-a^2 \mp 4a\sqrt{p}    }.
\label{varpieq}
\end{align}
We are now in a position to evaluate Eq. \eqref{Qe0recond} which becomes
\begin{align}
\frac{y_1}{y_2} = \frac{p(p-\varpi_+)}{ a^2 z_+^2}
=\frac{p-\varpi_+}{(p+2\varpi_+)} =\kappa.
\end{align}
Inserting the value for $\varpi_+$ from Eq. (\ref{varpieq}) leads to
\begin{align}
\kappa =  \frac{(p-6) p  -3 a^2   \pm   8 a \sqrt{p}}{p^2 +  3a^2\mp 4a \sqrt{p}  },
\label{kappaQe0}
\end{align}
where the upper (lower) sign corresponds to prograde (retrograde) orbits. This result can alternatively be obtained from the frequencies of linear perturbations to circular equatorial orbits \cite{1997ApJ...476..589P,2013LRR....16....1A}.
An equivalent way of expressing Eq. \eqref{kappaQe0}  is in terms of the quartic polynomial:
\begin{align}
\left[p (p-p^*)-a^2 (p^*-3)\right]^2-4 a^2 p (p^*-2)^2=0.
\label{polyp}
\end{align}
In the above expression we chose to use $p^*=6/(1-\kappa)$ to identify the resonance rather than~$\kappa$ itself. This choice makes it obvious that in the non-spinning limit $p=p^*$ is an analytic solution to the resonance condition.

Eq.~\eqref{polyp} is a key result of this paper because it characterizes the exact $'V'$ profile of all resonances for $z_- \to 0$ as a function of spin.  As discussed in Sec. \ref{pinfLimit}  on the weak field limit, eccentricity  has very little effect on the resonance surface and inclination merely deforms the $'V'$ profile into a line as $z_-\rightarrow 1$. This single formula thus allows us to characterize all resonant effects of arbitrarily spinning black holes.

To efficiently evaluate Eq.~\eqref{polyp} it is useful to view it as a quadratic polynomial in $a^2$ instead of a quartic polynomial in $p$. Solving for $a^2$ in terms of $p$ and $p^*$ leads to
\begin{align}
a_*^2(p)&=\frac{p^2 (p^*-3)+p
   \left(p^{*\ 2}-5 p^*+8\right)}{(p^*-3)^2} \notag\\
&+\frac{-2p(p^*-2) \sqrt{  p (p^*-3)-p^*+4}
}{(p^*-3)^2}.
\label{arp}
\end{align}
We use Eq. \eqref{arp} to plot the spin dependence of the $'V'$ profile for $z_-=0$ for several low-order resonances in Fig.~\ref{fig:spine0}.

The maximum splitting of the retrograde and prograde branches of the $'V'$ occurs when $a=1$ in this case the relevant roots of Eq.~\eqref{polyp} are,
\begin{align}
 p_{\pm} &= p^*-1\mp2\sqrt{p^*-2}.
\end{align}
The maximum opening distance of the $'V'$ profile is then
\begin{align}
(p_--p_+)= 4 \sqrt{p^*-2}.
\end{align}

Even though Eq. \eqref{polyp} can readily be solved for $p$, the expression is complicated and it is often difficult to identify which roots correspond to the retrograde and prograde branches, we thus provide a useful series expansion. For low spin values, the solutions to Eq. (\ref{polyp}) admit the expansion
\begin{align}
p_\mp=p^*\pm\frac{2 a (p^*-2)}{\sqrt{p^*}}-\frac{a^2 \left(p^{*2}-5
   p^*+8\right)}{p^{*2}} \notag\\
\pm\frac{a^3 (p^*-2) \left(2 p^{*2}-11
   p^*+20\right)}{p^{*7/2}}+O\left(a^4\right) \label{spinEQ}
.\end{align}
Table~\ref{TabRes} summarizes the numerical values associated with the low-order resonances depicted in Fig~\ref{fig:spine0}, both in dimensionless and physical units for the special case of the Galactic center, Sgr A*.  Lower order resonances, shown in bold in this table, are likely to have observationally detectable dynamics. According to the KAM theorem these tori are most likely to  be disrupted and the ensuing rapid changes in the orbital parameters should have a dramatic effect when compared to the systematic smooth distortion of perturbation induced effects away from resonant orbits. We will discuss this further in Sec.~\ref{astrophysics}. In Sec.~\ref{Numerical2o3}  we give a numerical characterization of several of the
lower order resonances introduced here. However before we turn to the numerical solution
we analytically quantify the effect of eccentricity in greater detail.

\subsection{Quantifying the effect of eccentricity}
To quantify the effect of eccentricity,
we consider the limiting case of polar orbits with
\begin{align}
z_-^2&=1, & L_z &= 0, &z_+^2=\frac{Q}{\beta^2}=\frac{p^2\varpi_{\times}}{a^4(1-e^2)}.
\end{align}
This choice puts us at the top of the inverted $'U'$ where the effects of spin are minimal.
In this case we can solve Eqs. \eqref{wptquad} and  \eqref{wpcond} for $\varpi_\times$ and $\varpi_-$. Since $L_z$ is zero Eq.~\eqref{wptquad} reduces to a linear equation. As a result, only one solution exists as is expected because the retrograde and prograde branches should coincide for polar orbits.

Here to illustrate the effect of eccentricity we only give the results in the $a=0$ limit, since the expressions for $a\neq 0$ are unwieldy, so that
\begin{align}
\varpi_\times&=0,& \varpi_+ &= \frac{2p}{p-4}.
\end{align}
Substituting these values into Eq. \eqref{paraExpand} gives expressions for the parameters that enter into the series expansion of Eq. \eqref{expansion},
\begin{align}
\frac{y_1}{y_2} &= \frac{p-6}{p },& \frac{\delta_1}{y_1} &= -\frac{2e}{p-6}, & \frac{\delta_2}{y_2}=0. \label{zm1a0}
\end{align}
We now re-expand the series in Eq. \eqref{expansion} and substitute Eq. \eqref{zm1a0} to obtain an explicit expression for $\kappa$,
\begin{align}
\kappa= (1-6/p) \left( 1-\frac{3 e^2}{2 (p-6)^2}-\frac{51 e^4}{32 (p-6)^4} + O(e^6)\right).
\notag
\end{align}
Inverting this series expansion we find that the dominant behavior of the resonance's dependence on eccentricity is,
\begin{align}
p=p^*\left(1+\frac{e^2}{4(p^*-6)} - \frac{e^4(4p^*-17)}{64(p^*-6)^3} + O(e^6)\right).
\label{pandecc}
\end{align}
The expansion given in Eq. \eqref{pandecc} is valid in the strong field region. Note that as the resonant surfaces approach the  ISCO ($p=6$) the effects of eccentricity become increasingly important.

\subsection{Detailed numerical characterization of low order resonances}
\label{Numerical2o3}
\begin{figure}
\includegraphics[width = \columnwidth]{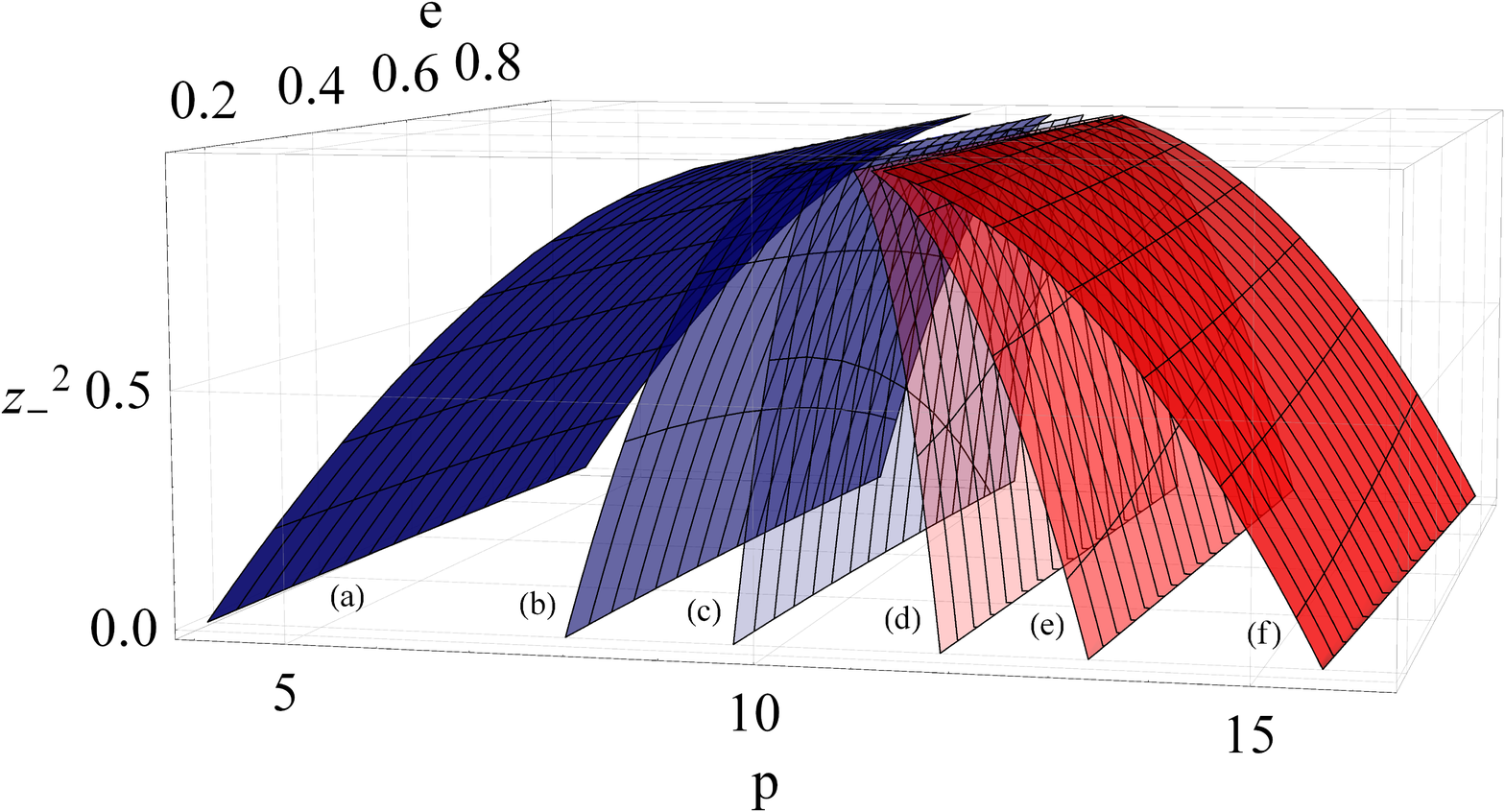}
\caption{\footnotesize{Numerically computed  surfaces for the 2/3 resonance. Prograde orbits are shown in blue (left) for spin values (a) $0.99$, (b) $0.5$ and (c) $0.2$. Retrograde orbits are shown in red (right) for spin values (d) $0.2$, (e) $0.5$ and (f) $0.99$.}}
\label{sheets}
\end{figure}

\begin{figure*}
\centering
\includegraphics[width = \textwidth]{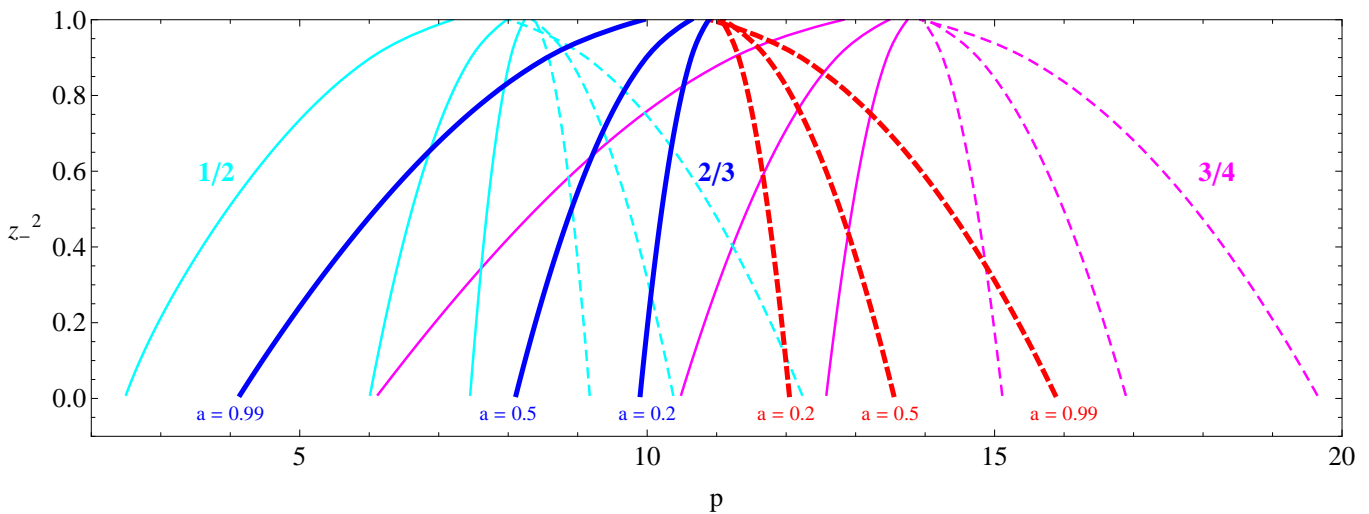}
\caption{
The numerically computed location of the $1/2$, $2/3$ and $3/4$ resonances for fixed eccentricity, $e = 0.6$ for the spin values $a=0.2, \ 0.5, \ 0.99$. (These are the same spin values used in Fig.~\ref{sheets}.) The resonances for prograde orbits are shown in solid lines and those for retrograde orbits in dashed lines. The resonances appear centered around the corresponding $p^{*}$ values given in Table~\ref{TabRes}, with the $1/2$-resonance (cyan) centered around $p^{*} = 8$,
the $2/3$ resonance (blue and red) centered around $p^{*} = 10.8$ and the $3/4$ resonances (magenta) around $p^{*} = 13.7$. The features of these plots agree qualitatively with those of 
Fig.~\ref{res0304eall}.
\label{122334}}
\end{figure*}

Having studied the resonances in limiting cases, 
we now provide a full numerical characterization of the properties of the low-order resonances and a comparison with the analytic formulas. We quantify the error in using the weak field approximation of  Sec.~\ref{pinfLimit} for low order resonant surfaces and confirm the veracity of the exact analytic solutions found for special cases in the strong field regime.

We calculated the location of the resonances numerically by means of two methods. 
The first method directly uses the closed-form general
expressions for the fundamental frequencies presented by Schmidt \cite{Schmidt} (see for
example Eqs. [33] and [34] where the frequencies are given in terms of elliptic
integrals).
These formulas are not well defined for the limiting cases
\mbox{$e = 0$} and \mbox{$z_{-} = 1$}  which require explicit modifications \cite{Schmidt}. The
integrals also become indeterminate when reducing to $a = 0$ or $a = 1$. 
A careful numerical treatment in the limiting cases $(e=0, \ z_-=0)$,
where the numerical values are calculated using the procedure presented in Appendix B
of \cite{Schmidt}, recovers the analytic values predicted by Eqs.~\eqref{arp} and \eqref{pandecc} to within $\lesssim 0.01 \%$ over the entire range
of the remaining parameters. The accuracy of the agreement in these cases are limited by
the numerical procedure, e.g. taking $a = 0.005$ and $z_-=0.99995$ instead of $0$ and
$1$ respectively.

The second semi-analytic scheme that we implemented exploits the analytic development of Sec. \eqref{EOM} and \eqref{Sec:resonanceCond} to reduce the computational cost in the following way. The resonance condition Eq.~\eqref{RfSpre} is recast in terms of the three parameters that appear in the  series expansion of Eq. \eqref{paraExpand} as
\begin{align}
\frac{\omega_r}{\omega_\theta} = \sqrt{\kappa}  =\sqrt{
\frac{y_1}{y_2 }}
\frac{R_F(0,1+\frac{\delta_2}{y_2} , 1-\frac{\delta_2}{y_2})  }{R_F(0,1+\frac{\delta_1}{y_1} , 1-\frac{\delta_1}{y_1})     }. \label{RfSpre2}
\end{align}
For a given $a,\  e,\  z_-^2>0$ we calculate the values of $p$ that lie on the resonance surfaces by first finding an expression for $\varpi_+$ in terms of $\varpi_\times$ using Eq. \eqref{wpcond} and substituting this into Eq.~\eqref{wptquad}.
Solving the resulting quadratic for $\varpi_\times$ yields two choices for $\varpi_\times(p)$, corresponding to the prograde and retrograde solutions. Having expressed $\varpi_\times$ and thus $\varpi_+$ in terms of $p$ we can write the three parameters in Eq.~\eqref{paraExpand}  as
explicit functions of $p$. The right-hand side of Eq.~\eqref{RfSpre2} can be evaluated for a given $p$ by repeatedly applying the relation \eqref{Rfxysq} which tends to make the two arguments in the $R_F(0,x^2,y^2)$ function  equal (5 iterations usually suffice to reach machine precision) and then using \eqref{Rsymm} to yield the result;  alternately one can use Eq.~\eqref{RFellip} to express Carlson's $R_F$ functions as elliptic integrals of the first kind. A line search method can then be used to find a value $p$ that solves \eqref{RfSpre2} for a given $\kappa$, using $p=p*$ as the first guess.
This method has  yielded accurate results for all parameter values except in the equatorial case $z_-=0$. The cases $e=0$, $z_-=1$, $a=0$ and $a=1$ do not require special treatment, in contrast to the first direct numerical method.
The special case of $z_-=0$ can be treated analytically, as was shown in the previous sections.

In the numerical investigation shown in Figs.~\ref{sheets} and \ref{122334}  we sample the parameter space
by choosing several values of $e$ and $z_-$ in the range $(0,1)$
and numerically solve the
equation \mbox{$\omega_r/\omega_\theta =1/2,\ 2/3\ \mbox{or}\ 3/4 $} for $p$, given a fixed spin value~$a$.
The root finding and function evaluations required to solve the implicit equation for $p$ are computationally expensive, particularly when using the first direct numerical method.
Rendering Fig.~\ref{sheets} in Mathematica
took two days on an Intel i7 PC with 8 cores  each with 2GB of RAM using the first method.
The second semi-analytic method that takes advantage of the commensurability of the frequencies and
is also implemented in Mathematica, yields the same result in just under an hour. (A further speed up is likely for a Matlab or C++ code). The results from the two methods agree and serve as independent checks.

The numerically computed resonant surfaces are plotted in Fig.~\ref{sheets}.  This figure shows several surfaces or sheets  corresponding to different black hole spins.
The sheets are plotted on the $\{e, p, z_{-}^2\}$ coordinate axis
to make the correspondence to Fig.~\ref{res0304eall} clear. The blue sheets $(a)$--$(c)$ indicate the location of prograde orbits.
The spin value  is decreased from sheet $(a)$ to $(c)$. The red sheets indicate retrograde orbits with spin values increasing from $(d)$ to $(f)$.
The higher the eccentricity, the further out in $p$, the resonances occur.  The more inclined the orbital plane, the further out (closer in) the resonances for prograde (retrograde) orbits occur. These qualitative features are the same as those identified in the weak field limit in Sec.~\ref{pinfLimit}.
Since the value of the orbital eccentricity
has the smallest impact on the location of the resonance, we
subsequently choose a representative eccentricity and explore the spin dependence for several low order resonances.
In Fig.~\ref{122334} the $1/2$, $2/3$ and $3/4$ resonances
   are shown for prograde and retrograde orbits with $e = 0.6$ around a Kerr black hole for several different
   spin values.
We see that the typical shape of the resonances shown in Fig.~\ref{sheets} is preserved. Prograde
   resonances with larger spin values are closer to the black hole in comparison to the point $p^{*}$ whereas retrograde resonances associated with higher spin
values occur further outward. The shapes of these resonances are in qualitative agreement with the weak field $'U'$-$'I'$ transition discussed with respect to Fig.~\ref{res0304eall}.

\begin{figure}
\includegraphics[width = \columnwidth]{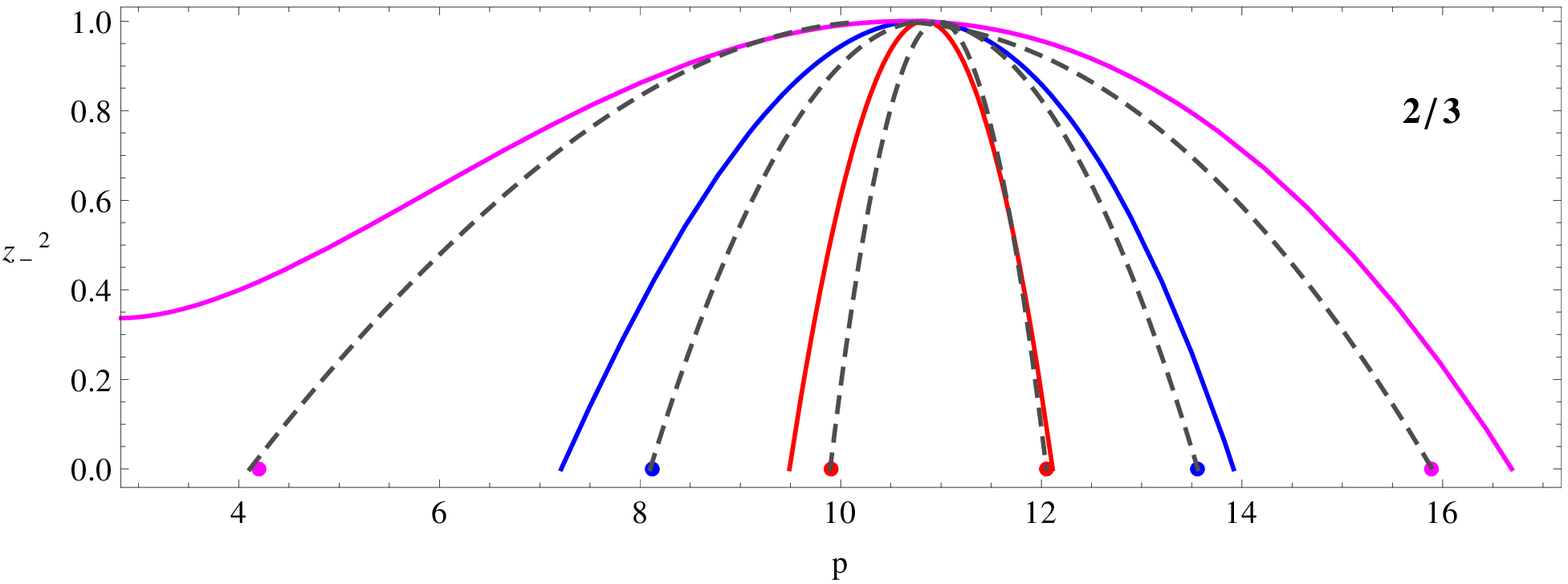}
\includegraphics[width = \columnwidth]{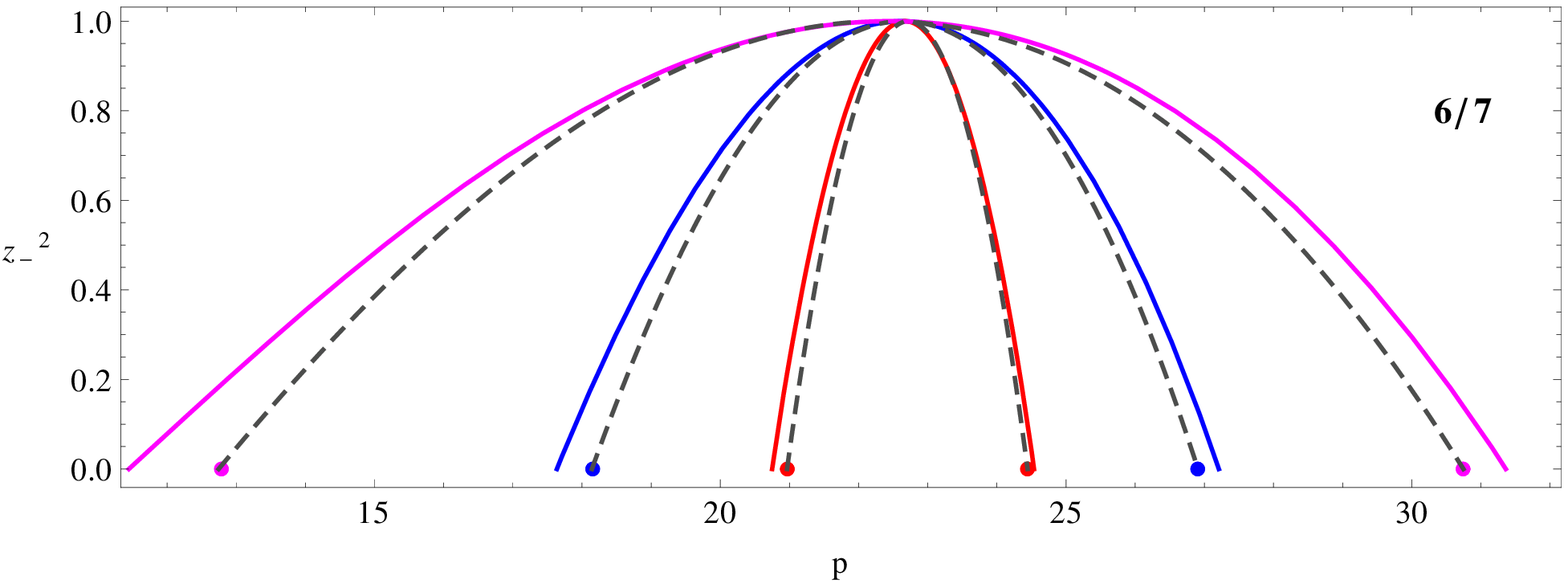}
\includegraphics[width = \columnwidth]{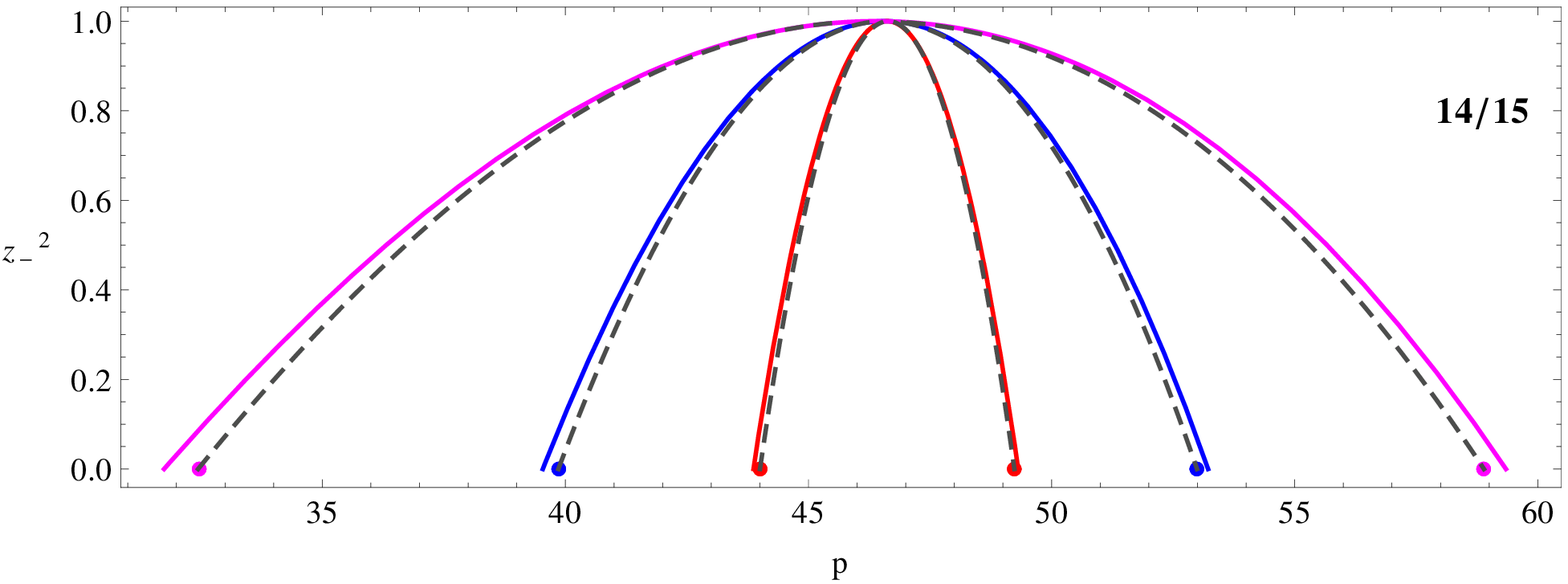}
\includegraphics[width = \columnwidth]{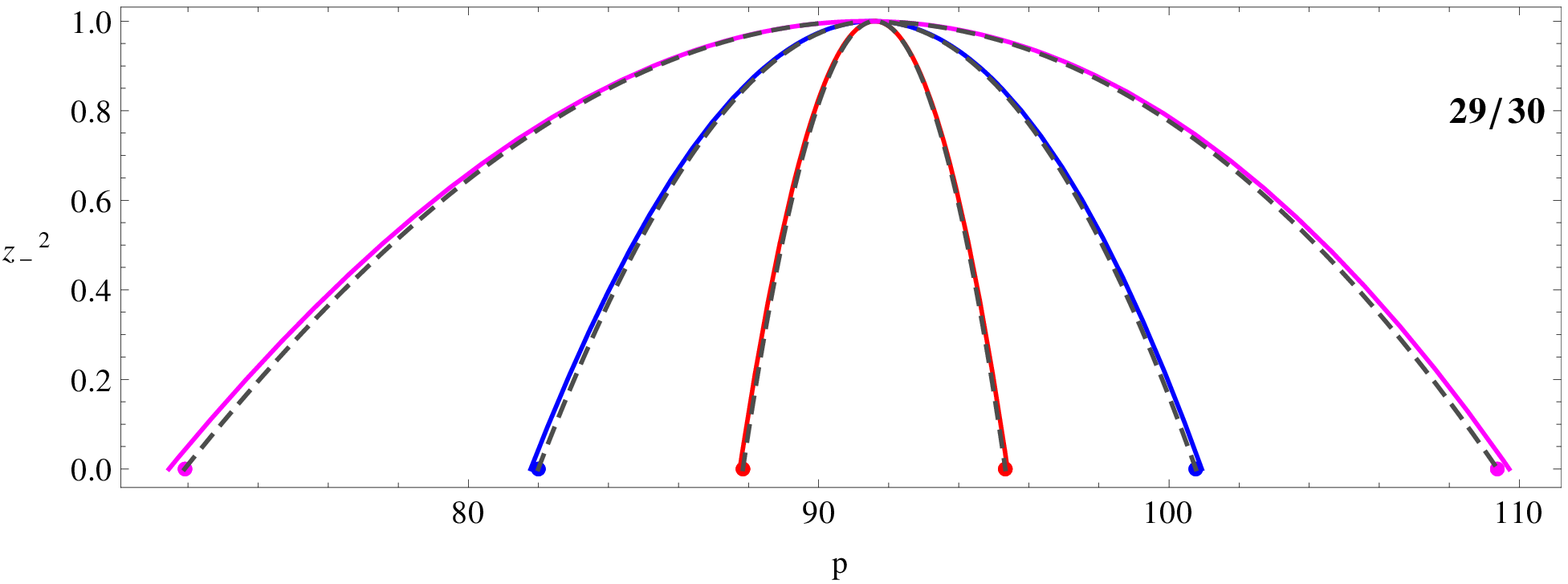}
\caption{\footnotesize{Quantitative comparison between the numerically calculated resonance surfaces  and weak field approximation of Sec.~\ref{pinfLimit} for the $2/3$, $6/7$, $14/15$ and $29/30$ 
resonances. The eccentricity is fixed at $e = 0.6$. The weak field approximation is computed using Eq. \eqref{z2meq} and is shown for spin values $a = 0.2$ in red (inner arch), $a=0.5$ in blue
(middle arch) and $a = 0.99$ in magenta (the outer arch). The analytic results on the equatorial plane are shown in matching colored dots, with the numerical solutions overlayed using black dashing.}}
\label{Numbrella}
\end{figure}

A quantitative comparison between the numerically computed position for the $2/3$, $6/7$, $14/15$ and $29/30$ resonance and the weak field approximation developed in Sec.~\ref{pinfLimit} is given in Fig. \ref{Numbrella}.
The analytic result is obtained from Eq. \eqref{z2meq}, with  $e = 0.6$ and varying spin values. The weak field approximation provides a  good fit for high orbital inclination ($z_{-}^2 = 1$) even when $p < 20$.
The  weak field approximation deviates more strongly from the analytic results as the spin value increases, such that the orbits with $a= 0.2$ exhibit a better fit than those for which $a = 0.99$. For a given spin value the approximation fits the retrograde orbits (higher $p$) better than the corresponding prograde orbits (lower $p$). This is to be expected since in the approximation of Sec.~\ref{pinfLimit} we used a quadratic approximation to the resonance
condition valid only for $p\to \infty$ and included spin effects only up to $O(a^2/p^2)$. Fig. \ref{Numbrella} also shows the analytic solution on the equatorial plane as calculated from the roots of Eq.\eqref{polyp} with the added second order correction for non-zero eccentricity $p^{*}e^2/4(p^{*}-6)$ given in Eq.~\eqref{pandecc}. These equatorial solutions are in very good agreement with the numerical results. They provide an easily computed indicator of where the weak field approximation strongly departs from the analytically computed surfaces.

\section{The location of Resonant surfaces terms of $E$, $L_z$ parameters}
\label{Sec:EL}

When conducting numerical investigations of orbits in spacetimes that are more general than Kerr it is often useful to plot a Poincar\'{e} map at fixed
$E$ and $L_z$ as a diagnostic tool to explore the break down of KAM-tori.  This method of surveying the parameter space is not as well suited for describing the physical location of the orbit, but it  unambiguously generalizes to metrics where a Carter constant $Q$ does not exist. In this section we develop some intuition for the features of resonant surfaces in $E$, $L_z$ space.  This picture will be invaluable as we discuss
departures from integrability and torus breakdown in the next section.

\begin{figure}[h]
\includegraphics[width = \columnwidth]{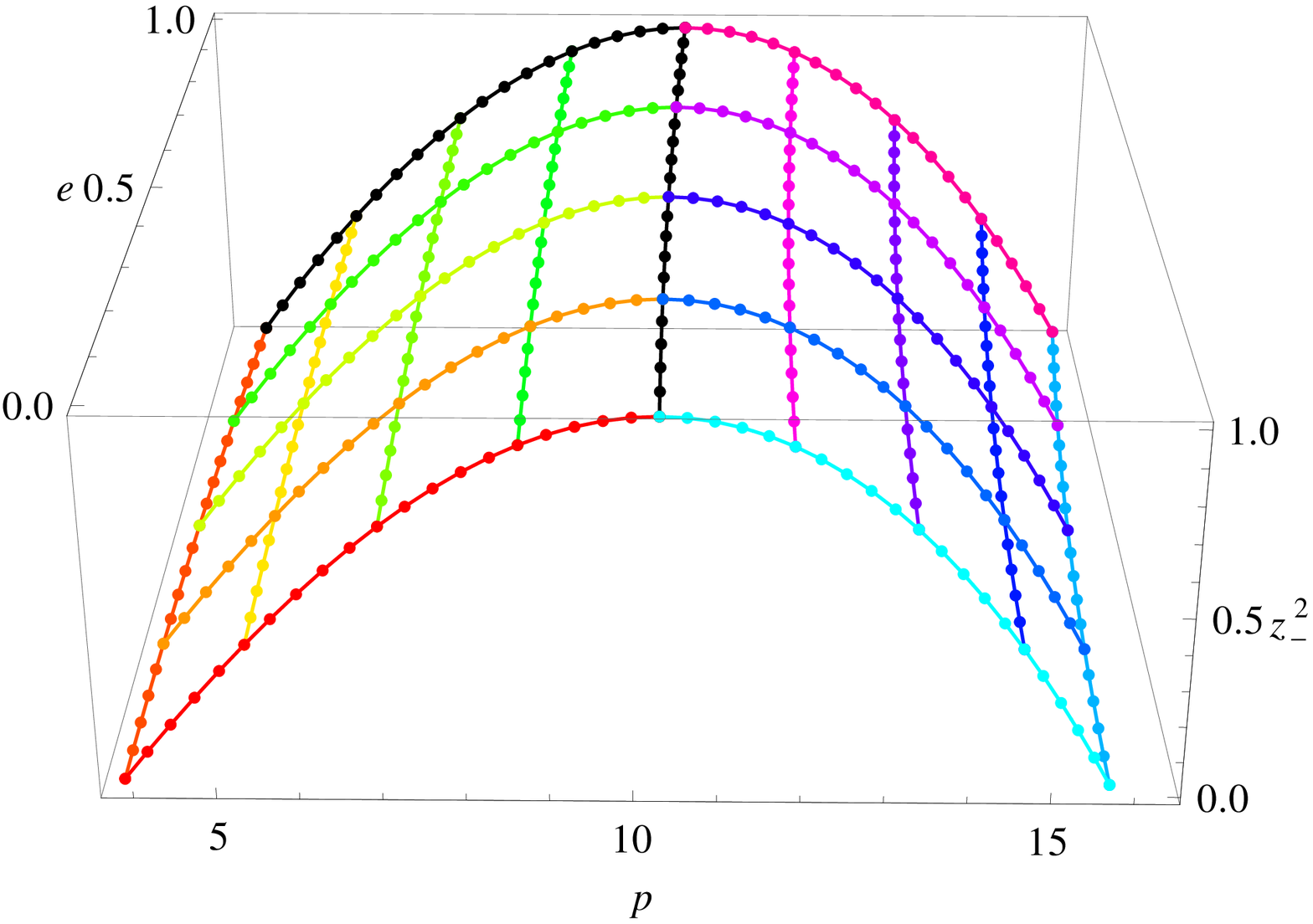}
\caption{The  $2/3$ resonant surface for a maximally spinning black hole $a=1$ represented in terms of $(p, e, z_-^2)$ coordinates.
 Horizontal lines indicate longitudinal values of $z_-^2 =\{ 0, \sin\frac{\pi}{8}, 1/\sqrt{2}, \cos \frac{\pi}{8},1\}$, the bent arches indicated eccentricities of $e=\{0,\frac{1}{4},\frac{1}{2}, \frac{3}{4}, 1\}$
\label{eplELZ}}
\includegraphics[width = \columnwidth]{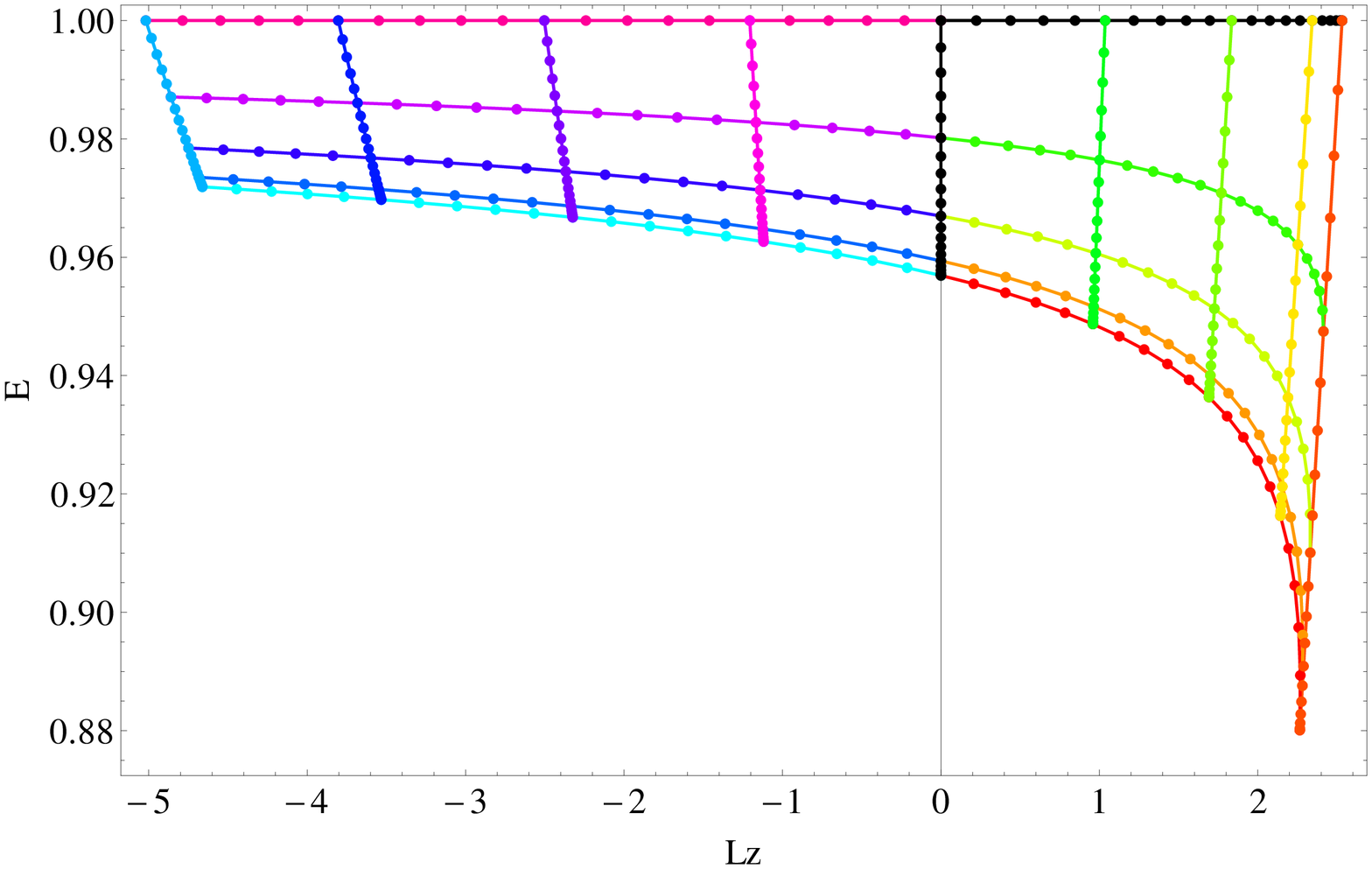}
\caption{The $2/3$ resonant surface for  $a=1$ projected onto $E$, $L_z$ coordinates \cite{ourPRL}.
 Lines indicate constant values of \mbox{$z_-^2 =\{ 0, \sin\frac{\pi}{8}, 1/\sqrt{2}, \cos\frac{\pi}{8},1\}$}, and \mbox{$e=\{0,\frac{1}{4},\frac{1}{2}, \frac{3}{4}, 1\}$} corresponding to those in Fig.~\ref{eplELZ}. The large asymmetry across the $L_z = 0$ line is indicative of the large spin value under consideration.
\label{eplELZ2}}

\end{figure}

\begin{figure}
\includegraphics[width = \columnwidth]{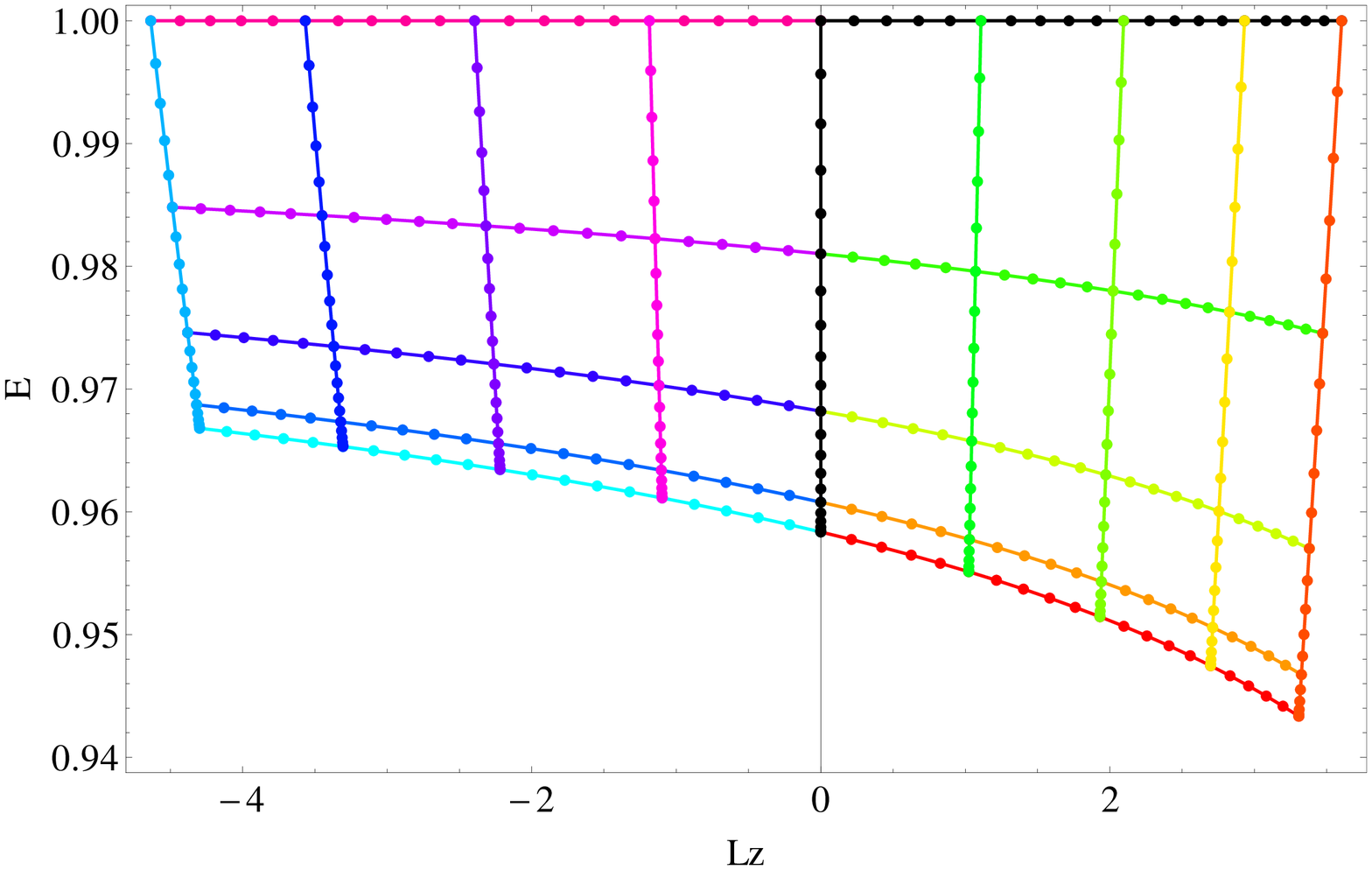}
\caption{The $2/3$ resonance for  $a=0.5$ in $E$, $L_z$ coordinates for
   \mbox{$z_-^2 =\{ 0, \sin\frac{\pi}{8}, 1/\sqrt{2}, \cos\frac{\pi}{8},1\}$}, and \mbox{$e=\{0,\frac{1}{4},\frac{1}{2}, \frac{3}{4}, 1\}$}. Note the decrease asymmetry across  $L_z = 0$ line when compared to Fig. \ref{eplELZ2} which is maximally spinning.
\label{eplELZ2ap5}}
\includegraphics[width = \columnwidth]{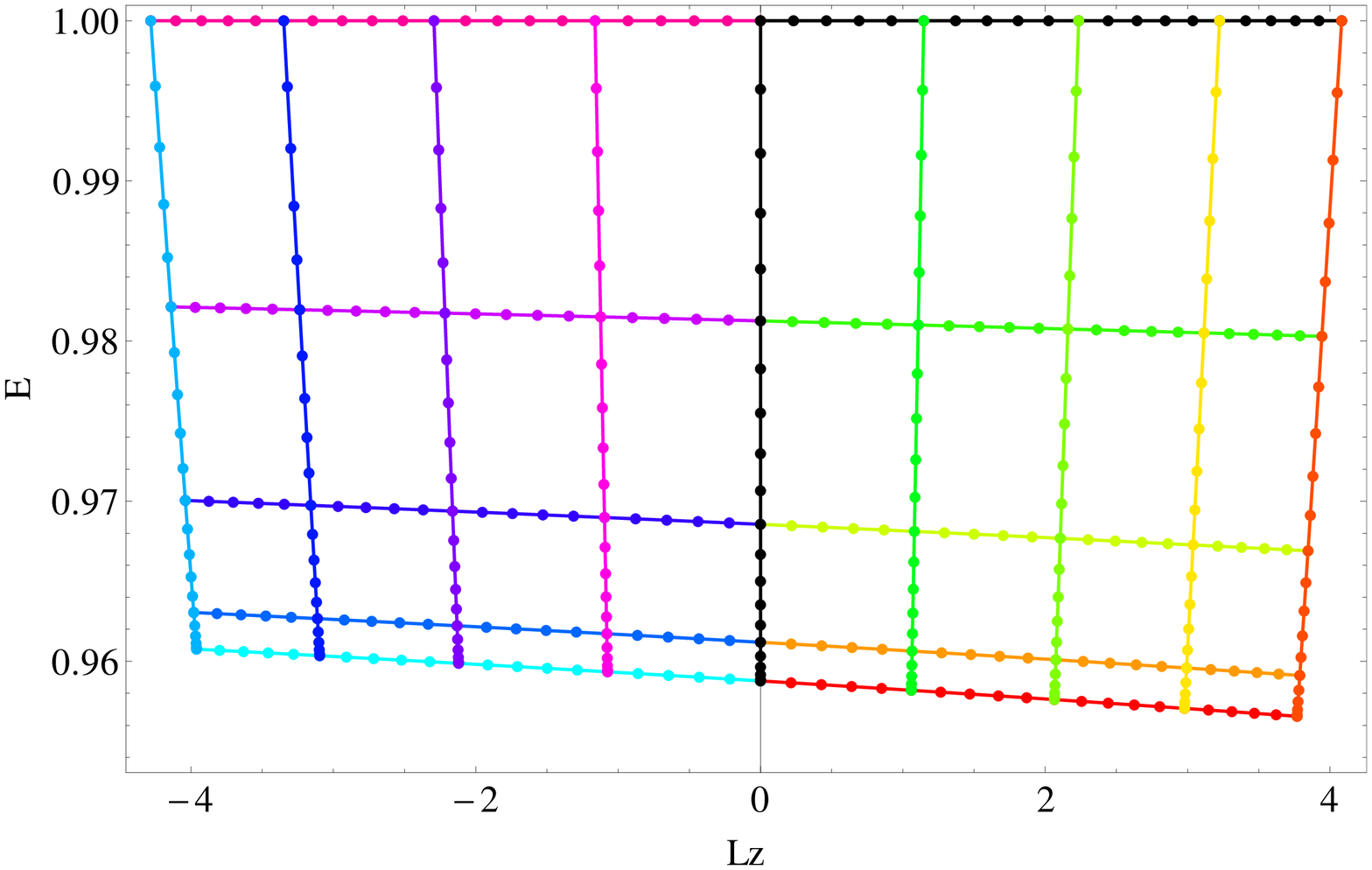}
\caption{The $2/3$ resonance for  $a=0.1$ in $E$, $L_z$ coordinates for
  \mbox{$z_-^2 =\{ 0, \sin\frac{\pi}{8}, 1/\sqrt{2}, \cos\frac{\pi}{8},1\}$}, and \mbox{$e=\{0,\frac{1}{4},\frac{1}{2}, \frac{3}{4}, 1\}$}. Since the spin is almost vanishing the figure is much more symmetrical about the $L_z$  axis than  Fig.\ref{eplELZ2ap5} and \ref{eplELZ2}.
\label{eplELZ2ap1}}
\includegraphics[width = \columnwidth]{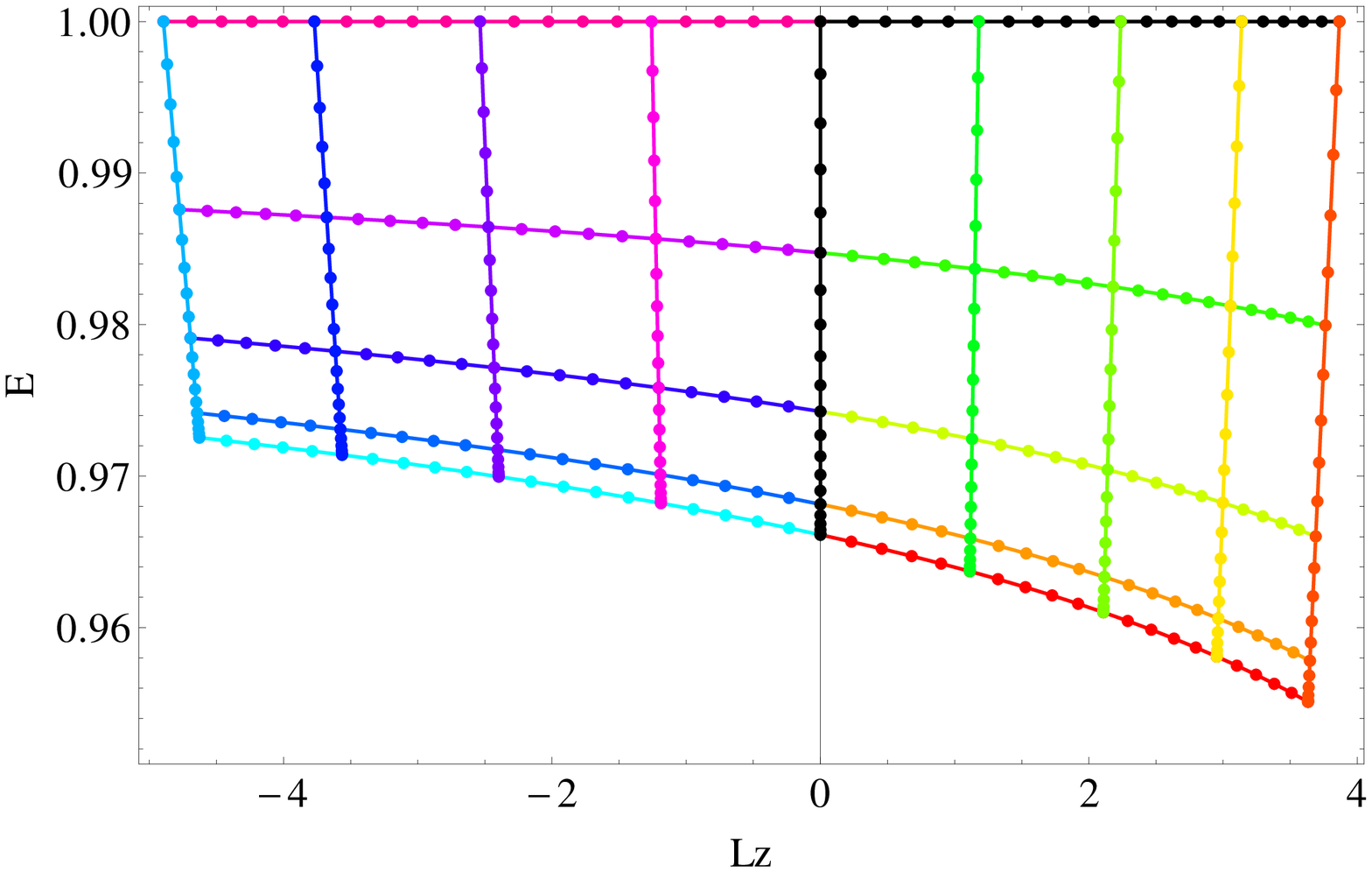}
\caption{The $3/4$ resonant surface for  $a=0.5$ projected onto $E$, $L_z$ coordinates, for \mbox{$z_-^2 =\{ 0, \sin\frac{\pi}{8}, 1/\sqrt{2}, \cos\frac{\pi}{8},1\}$}, and \mbox{$e=\{0,\frac{1}{4},\frac{1}{2}, \frac{3}{4}, 1\}$}.
The main difference between the parameter space covered by the $2/3$ and $3/4$ resonance in Fig.\ref{eplELZ2ap5} is that slightly lower energies are reached for the lower order resonance and for the higher order resonance the $L_z$ values have a marginally larger span, the other features remain qualitatively similar.
\label{eplELZ2ap5r34}}
\end{figure}

In Fig.~\ref{eplELZ} we show the $2/3$ resonant surface for a maximally spinning black-hole, $a=1$, computed using the second numerical method of Sec. \ref{Numerical2o3}. In this figure the dots correspond to computed points on the resonance surface.
The lines indicate fixed values of eccentricity and longitudinal angles at which the sampling took place: horizontal lines correspond to $z_-^2 =\{ 0, \sin(\pi/8), 1/\sqrt{2}, \cos(\pi/8),1\}$, the bent arches correspond to $e=\{0,1/4,1/2, 3/4, 1\}$. For each point ($p$, $e$, $z_-^2$) on the resonant surface, the corresponding $E$, and $L_z$ values were computed using Eqs.~\eqref{condQEL} and \eqref{Condelz}. Care was taken to compute the appropriate $\varpi_\times$ root associated with the retrograde $p_-$ or prograde $p_+$ branch of the resonant surface under consideration. The outcome is shown in Fig.~\ref{eplELZ2}. Note the large asymmetry about the $L_z=0$ line that is due to the fact that the black hole is maximally spinning. The low energy spike that coincides with prograde orbits close to the equatorial plane with low eccentricity is easy to miss numerically when exploring the parameter space in terms of $E$, $L_z$ parameters.
This spike constitutes a set of parameter values that potentially have important astrophysical implications: Most particles in thin astrophysical discs around rapidly spinning black holes are expected to be prograde and on nearly circular equatorial orbits.  As the spin of the black-hole decreases the resonance footprint on the $E$, $L_z$ plane will become increasingly symmetrical as is shown in Figs.~\ref{eplELZ2ap5} and \ref{eplELZ2ap1}

In Fig. \ref{eplELZ2ap5r34} the $E$, $L_z$ projection of the $3/4$ resonance surface is given for a black-hole with spin $a=0.5$. The higher order resonance surface has  features that are qualitatively similar to that of  the $2/3$ resonance of equal spin shown in Fig. \ref{eplELZ2ap5}. For the higher order resonances however the $E$ values are slightly higher and the $L_z$ values span a marginally larger range. 

From the figures in this section it should be clear that the representation of orbital parameters in terms of the $(p, e, z_-^2)$ coordinates is better adapted to the geometry of the resonant orbit, making it generally easier to quantify the resonance surfaces using these variables.  Once obtained the resonant surfaces projected onto the $E$, and $L_z$ orbital variables can potentially be used to aid numerical exploration into Kerr-like spacetimes discussed in the next section.

\section{Rotation Curves and the breakdown of KAM tori  }
\label{Sec:ROT}
\begin{figure}
\includegraphics[width = \columnwidth]{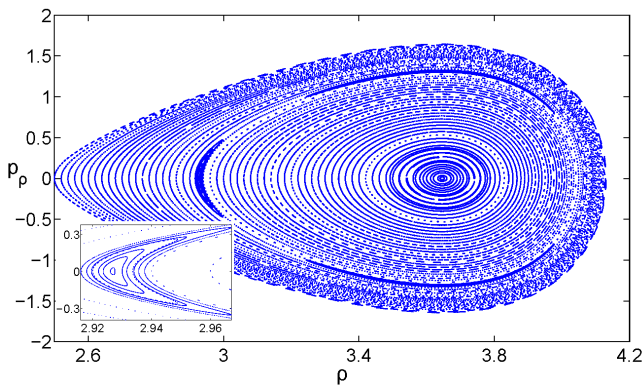}
\caption{A Poincar\'{e} map generated for the Manko-Novikov metric.  The inlay shows a close-up of one of the islands in the Birkhoff chain of multiplicity three. This island is associated with the breakdown of the $2/3$ resonance as indicated using the rotation curve in Fig.\ref{rotationNo}
The spin value for  this simulation  was chosen to be $a = 0.9$ and the dimensionless quadrupole deviation parameter $q = 0.95$ was chosen to be very large. The other orbital parameters are $E = 0.95$, $L_{z} = 3$ and rest mass $\mu = 1$.
\label{MankoPoin}}
\end{figure}

A number of groups have numerically explored the breakdown of integrability in stationary axisymmetric vacuum metrics such as the Manko-Novikov metric that reduces to the Kerr metric for a certain choice of parameters \cite{Kostas,Contopoulos2012,Gair, Contopoulos2011, MGeyerThesis}. 
One of the tools used to explore the non-Kerr spacetimes numerically is to plot the Poincar\'{e} map of the orbital motion for a
fixed choice of energy ($E$) and angular momentum ($L_z$). An example of a Poincar\'{e} map exhibiting broken tori is given in Fig. \ref{MankoPoin}.
Each closed curve in the Poincar\'{e} map corresponds to an orbit that was started on the equatorial plane and given an initial momentum out of the plane. Each intersection of the orbit with the equatorial plane generates a point on the Poincar\'{e} map. The radial coordinate (here $\rho$) and corresponding radial momentum ($p_{\rho}$) of these piercing points are plotted. The closed curves in this Poincar\'{e} map indicate that for most initial conditions geodesic motion in the Manko-Novikov metric remains integrable or regular. However, among the closed curves, a Birkhoff chain of islands can be seen (enlarged in inset), which indicates that for certain initial conditions the regularity of orbits break down. This breaking of regular structure is associated with resonances in the fundamental frequencies describing the orbit as predicted by the KAM theorem.

\begin{figure}
\includegraphics[width = \columnwidth]{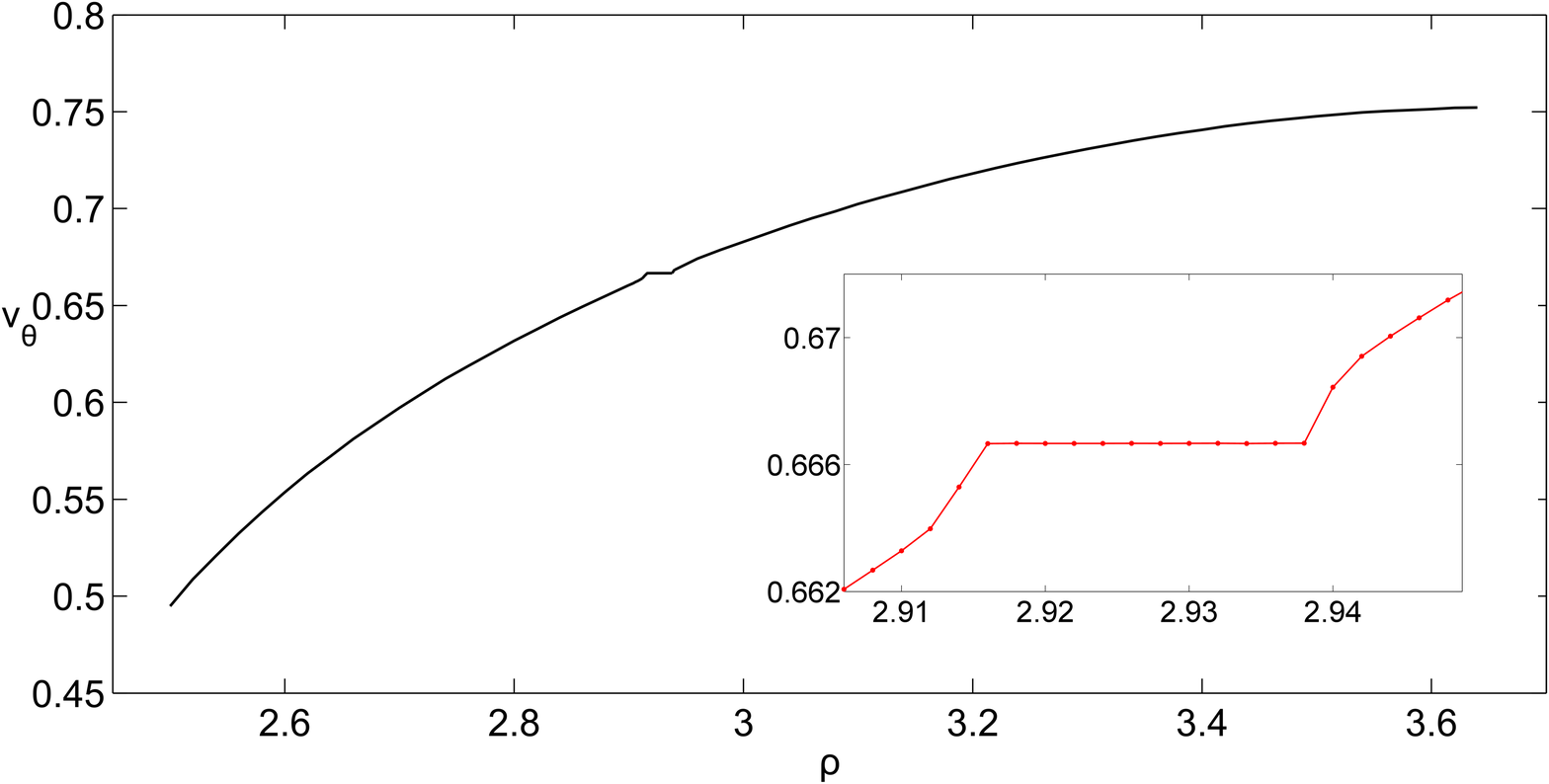}
\caption{\footnotesize{The rotation curve is obtained by calculating the rotation number of each orbit from the Poincar\'{e} map given in Figure~\ref{MankoPoin}. The rotation number, $\nu_{\theta}$, which is equal to the ratio of orbital frequencies, $\omega_{r}/\omega_{\theta}$, is plotted as a function of the initial radial coordinate of the orbit. The rotation curve is generally smooth and monotonically increasing. However, corresponding to the Birkhoff chains of islands seen in the Poincar\'{e} map, there is a plateau in the rotation curve for which $\omega_{r}/\omega_{\theta} = 2/3$ is resonant.}}
\label{rotationNo}
\end{figure}

The second diagnostic  tool that gives insight during a numerical exploration is the rotation curve. This is a plot
of the $\omega_r/\omega_\theta$ frequency as a function of initial conditions for the Hamiltonian potential being studied. For an integrable system the rotation curve is a smooth function of initial conditions.  In systems for which the KAM tori have broken, there are plateaus in the rotation curve where the numerically computed rotation number \cite{Contopoulosbook,Voglis1998} remains roughly constant. This situation is depicted in Fig.~\ref{rotationNo}.

In the numerical investigation of the Manko-Novikov metric,
 the breaking of low-order resonant tori is observed most dramatically
at the $2/3$ resonance. The apparent dominance of the $2/3$ resonance over other integer ratios could heuristically be explained by the fact that in these studies, the deviation from Kerr was mainly a quadrupole perturbation and thus roughly proportional to $\cos 2\theta$. This geometric dependence of the perturbation indicates that any $2/m$ resonance is expected to be strongly excited. Integers other than $n=2$ would arise from the nonlinear coupling between the angle variables and the coordinates, which is a higher order effect. Since $\omega_r<\omega_\theta$ for Kerr we always have that $m>n$, so $2/3$ is the resonance with the  lowest order integer ratio. It is thus expected to dominate the breakdown of integrability for a system subjected to  a quadrupole perturbation.

 The numerical explorations of the breakdown of KAM tori has to date been limited to a very small subset of the allowed parameter space.  
To guide future studies covering the entire parameter space associated with the $2/3$ and $3/4$ resonances discussed in Sec. \ref{Sec:EL},
we now describe how to use the machinery developed in this paper to  analytically compute the rotation curve for the Kerr metric.
An important caveat to bear in mind when exploring large deviations from
the Kerr metric is that although the  torus structure is not destroyed it may be distorted and the breakdown of quasi-periodic orbits may be shifted from the exact location of the resonance predicted by the rotation curve for orbits in Kerr spacetime. To zeroth order, however,  the Kerr rotation curve  provides an indication of the interesting regions in the parameter space.

\begin{figure}
\includegraphics[width = \columnwidth]{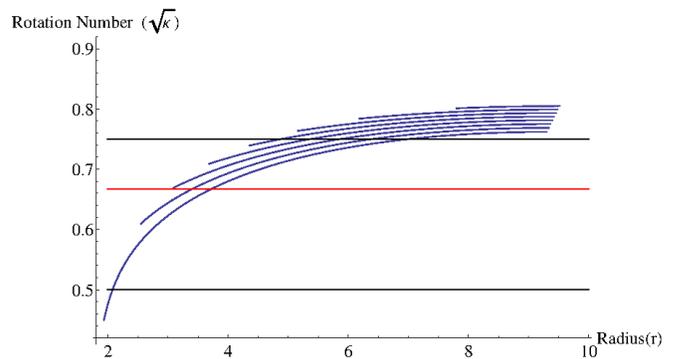}
\caption{The rotation curves for geodesics with  $E=0.95\mu$,  in a Kerr black hole spacetime with spin $a= 9/10$. The different curves correspond to different $L_z$ values uniformly spaced between $L_z=2.482\mu$ (lower curve) and $L_z=3.343 \mu$ (upper curve). For reference purposes, the 2/3 resonance value is indicated by the central red-line, the 1/2 resonance by the lower black-line and the 3/4 resonance by the upper black-line.  The radius (r) refers to the radius of closest approach of the orbit to the black hole \label{KerrROTC}}
\end{figure}

For the Kerr metric the rotation curve is computed by evaluating Eq.~\eqref{RfSpre2}
for a fixed $E$ and $L_z$ using the parameters defined in  Eq. \eqref{paraExpand}. 
A representative example is given in Fig.~\ref{KerrROTC}.
For a given $E$ and $L_z$, bound orbital motion is usually restricted to a small region of physical space.  Equatorial orbits ($Q=z_-=0$) have the largest radial extent while circular orbits ($e=0$) have the largest longitudinal reach, $z_{-\text{max}} = z_-(e=0)$. These two extreme points define the boundaries of the rotation curve. We examine them more closely by first computing the variables $\{p, e, z_-\}$ and subsequently the ratio $\omega_r/ \omega_ \theta$.

On the equatorial plane, $z_{- {\rm eq}}=0$ and the radial potential Eq.~\eqref{Rpot}  has one zero root, thus $\varpi_\times=0$. The largest two roots of the remaining cubic determines $p_{{\rm eq}}$ and $e_{{\rm eq}}$.
First consider the effect of  eccentricity $e_{{\rm eq}}$ where the  two extreme cases for bound orbits are $e_{{\rm eq}}=0$  and $e_{{\rm eq}}=1$.
Making use of Eq.~\eqref{kappaQe0} we have that if $e_{{\rm eq}}=0$ the rotation curve reduces to single dot at
\begin{align}
\omega_r/\omega_\theta = \sqrt{  \frac{(p-6) p  -3 a^2   \pm   8 a \sqrt{p}}{p^2 +  3a^2\mp 4a \sqrt{p}  }   }. \label{spindot}
\end{align}
If the $E$ and $L_z$ values have been chosen to lie on the $2/3$ resonance,
this value would correspond to  either the lower rightmost or lower leftmost points in Figs. \ref{eplELZ2},  \ref{eplELZ2ap5} and \ref{eplELZ2ap1} depending on the spin of the black-hole and whether the retrograde (left) or prograde (right) orbit is under consideration.
Eq.~\eqref{spindot} in conjunction with Eqs.~\eqref{condQELQ0} and~\eqref{varpieq}
can be used to estimate the characteristic $E$ and $L_z$ associated with the $\kappa$ resonance in the non-spinning limit to be
\begin{align}
\frac{E^2}{\mu^2}&= \frac{2(2+\kappa)^2}{9(1+\kappa)}, & \frac{L_z^2}{\mu^2} &= \frac{12  }{1-
\kappa  }.
\end{align}
For the $2/3$ resonance this evaluates to
 $E = 0.95879 \mu$ and $L_z = \pm3.86702 \mu$.
If the spin is maximal, $a=1$,  the prograde branch has $E=0.88009 \mu$, $L_z =2.26471 \mu$, $p = 3.86704$ while the retrograde branch has
 $E=0.971899 \mu$, $L_z =-4.6614 \mu$, $p = 15.733$. These values can be compared to the lower corner points in Fig.~\ref{eplELZ2}.

For parabolic equatorial orbits,  $e_{{\rm eq}} \rightarrow 1$ and $E=\mu$. This case  corresponds to the largest possible radial extent a marginally bound orbit can have. For the $2/3$ resonance these values are indicated by the upper leftmost and upper rightmost points on Figs.  \ref{eplELZ2},  \ref{eplELZ2ap5} and \ref{eplELZ2ap1}.
We define the dimensionless angular momentum parameter $l_z=L_z/\mu$ and note that on the equatorial plane, $ \varpi_\times =0$,  $\varpi_+ = l_z/2$. Using Eqs.~\eqref{condQEL} and~\eqref{Condelz} implies that the values of $p$ are restricted to
\begin{align}
p &= \frac{l_z^2}{2}\left(1 \pm \sqrt{
\left(1 - \frac{4}{l_z}+\frac{4a}{l_z^2} \right)
\left(1 + \frac{4}{l_z}-\frac{4a}{l_z^2} \right)} \right).
\label{peqecc1}
\end{align}
The factor under the square root is positive only if \mbox{$l_z>2(1+\sqrt{1-a   })$} or $l_z< -2 (1+\sqrt{1+a}    )  $.
The smallest value of $p$ that can be attained for equatorial parabolic orbits is thus \mbox{$p_{\pm}= 2(1+\sqrt{1 \mp a})^2$} for prograde ($p_+$) and retrograde ($p_-$) orbits respectively.

The parameters that enter into Eq.~\eqref{expansion} for equatorial parabolic orbits are
\begin{align}
\frac{y_1}{y_2}  &= \frac{3p  -l_z^2}{  2 l_z^2 },  &
\frac{\delta_1}{y_1} &= \frac{
 l_z^2 -p  }{ l_z^2  -3p }, & \frac{\delta_2}{y_2}&=0.
\label{paramEQecc1}
\end{align}
Evaluating Eq.~\eqref{RfSpre2}
we find that $\sqrt{\kappa}$ is a monotonically increasing function of $p$ or equivalently $l_z$. In the limit of a spin zero black hole the resonance $\sqrt{\kappa}=2/3$ occurs when \mbox{$p=11.3273$}, \mbox{$l_z=4.18461$}. Parabolic orbits with larger $|l_z|$ will not sample the $2/3$ resonance regardless of inclination. For a maximally spinning black hole, the $2/3$ resonance occurs
at \mbox{$p_- = 16.2914$}, \mbox{$l_z=-5.01845$} for retrograde orbits and at $p_+=4.14634$, $l_z=2.53177$ for prograde orbits (see  Fig.~\ref{eplELZ2}).

Finally we consider the maximal vertical extent an orbit can have. This occurs for circular orbits given a fixed $E$ and $L_z$. If $e=0$, $\delta_1/y_1 =0$ and
\mbox{$\varpi_+=-2p +\frac{2 }{1-E^2/\mu^2} $}. Eq.~\eqref{Condelz} gives $\varpi_\times$ to be
\begin{align}
\varpi_\times &= \frac{a^2 \left(1-2 \frac{E^2}{\mu^2}\right)+2 a
   \frac{E}{\mu} l_z+p \left(\frac{E^2}{\mu^2} (p-3) p-
   (p-2)^2\right)}{(p-1) \left(\frac{E^2}{\mu^2}-1\right)},
\end{align}
which can be substituted back into $L_z^2$ in Eq. \eqref{condQEL} to give a polynomial in  $p$ that can be solved to find $p_{e=0}$ for the given $E$ and  $L_z$.
Substituting these results into
the remaining expansion parameters of Eq. \eqref{paraExpand} and evaluating
 Eq.~\eqref{RfSpre2}  yields the maximum value the rotation curve attains.

\section{Estimating the size of the perturbation that results in a dramatic change of dynamics due to Torus break down}
\label{Breakdown}
The KAM criterion for the possible destruction of tori can be augmented by further estimates. 
A generic analytic perturbation $F(r, \theta, p_r, p_\theta, p_\phi)$ to geodesic quantities such as the energy can be written as a Fourier expansion of the form
\be
F(r, \theta, p_r, p_\theta, p_\phi)=\sum_{\bfn=-\infty}^\infty F_\bfn (e, p, z_-)e^{i \bfn \cdot \bfq}, \label{eq:FFourier}
\ee
where $\bfq$ are the angle variables corresponding to the frequencies via 
\be
\frac{d \mathbf{q}}{d\lambda}=\bfomega+O(\epsilon) . \label{eq:angles}
\ee
For most orbits, all the Fourier components in Eq. (\ref{eq:FFourier}) with $\bfn\neq 0$ will be oscillatory and their contribution approximately averages out over an orbit. The orbit's leading order secular evolution is driven by the component with $\bfn=0$. However, at a resonance where $\bfk\cdot \bfomega=0$ the Fourier components in Eq. (\ref{eq:FFourier}) that involve the resonant combination $(\bfk \cdot \bfq)$ momentarily cease to be oscillatory since their phase becomes stationary. These components thus generically contribute order unity corrections to the secular evolution over the resonance time, except when the amplitude of these components is $O(\epsilon)$ or smaller. Since for analytic functions the Fourier amplitudes $F_\bfk$ fall off exponentially with $\bfk$, we expect the resonance to have an appreciable effect on the dynamics only when
\be
O_\bfk =\sum_i|k_i| \lesssim O(| \ln (\epsilon)|) . \label{eq:analyticFF}\ee  This criterion on $\bfk$ in (\ref{eq:analyticFF}) for ``essential'' resonances \cite{1988macc.book.....A} refines the  bound for sufficient irrationality discussed in Eq. (\ref{eqKAM}) for the special case of analytic perturbations.

For EMRIs with the only source of perturbation being the gravitational self-force, even the low-order resonances are expected to be weak in the sense that the dominant dissipation $F_\bfzero$ is generically larger than the resonance potential related to $F_\bfk$. The resonances will therefore appreciably modify but not destroy the object's continued in-spiral.  Now, however, consider the case of additional perturbations that lead to very strong resonance modifications, where heuristically $F_{\bfk}\gg F_{\bfzero}$. Chirikov's resonance overlap criterion  \cite{1979PhR....52..263C} specifies the conditions under which the complete loss of quasi-periodic motion is expected to occur as follows. Each strong resonance captures orbits in its vicinity and is thus surrounded by an oscillation region similar to the Birkhoff islands shown in Fig. (\ref{MankoPoin}) and manifests as a plateau in the frequency evolution similar to that in Fig (\ref{rotationNo}). The onset of full-blown chaos occurs when 
these Birkhoff chains associated with nearby resonances become large enough that they touch. More precisely, Chirikov's criterion states that the transition to stochastic behavior arises when two neighboring strong resonances overlap in the sense that the frequency width of their oscillation regions is larger than their spacing in frequency. An estimate for the width of the resonance regions is \cite{1979PhR....52..263C,BHprep}
\be
\Delta |\bfk\cdot \bfomega|\sim \sqrt{\epsilon}\sqrt{|(\bfk \cdot \bfomega)_{, J_\alpha}
\, F_{\alpha \bfk}|},  \label{eq:resonancewidth} \ee
where $J_\alpha=(e, p, z_-)$ and $F_{\alpha \bfk}$ are the forces such that $\dot J_{\alpha}=\epsilon F_\alpha+O(\epsilon^2)$ and $F_\alpha$ is Fourier expanded as explained above.
 Overlap occurs for two resonances associated with lattice vectors $\bfk$ and $\bfk^\prime$ respectively when
\be
\Delta \omega^{(\bfk)}_i+\Delta \omega^{(\bfk^\prime)}_i > |\omega^{\bfk}_i-\omega^{\bfk^\prime}_i|
\ee
  for each of the frequencies $\omega_i$ evaluated at the resonances of the unperturbed system. This estimate of the criterion for overlap is only a crude indicator for the transition to stochasticity, and in cases of interest the local dynamics must be systematically analyzed  \cite{1981JSP....26..257E, 2002PhR...365....1C}.

For most systems explored numerically to date, the plateaus in the rotation number curve remain small, and thus do not satisfy Chirikov's criterion. This indicates that we expect weak chaos in the sense that torus disruption is limited to a small region in phase space.

\section{Astrophysical implications}
\label{astrophysics} 
For astrophysical processes near
super-massive black holes
the Kerr geometry usually provides
the background stage on which several small perturbation
effects  play their part.
In select regions of the spacetime corresponding to low-order resonant orbits 
the smooth distortion due to perturbation
induced effects is disrupted. The background spacetime geometry largely sets the location of these resonance induced disruptions imprinted on the dynamical structures of the perturbed system,
 while their details depend on the properties of the perturbations.

Generally these effects are expected to be small, although persistent and robust.
They are induced by all non-integrable perturbations of the Kerr metric. If the perturbation is not too large resonant effects occur at the locations quantified in this paper irrespective of the
source of the perturbation.
\begin{figure}
\includegraphics[width = \columnwidth]{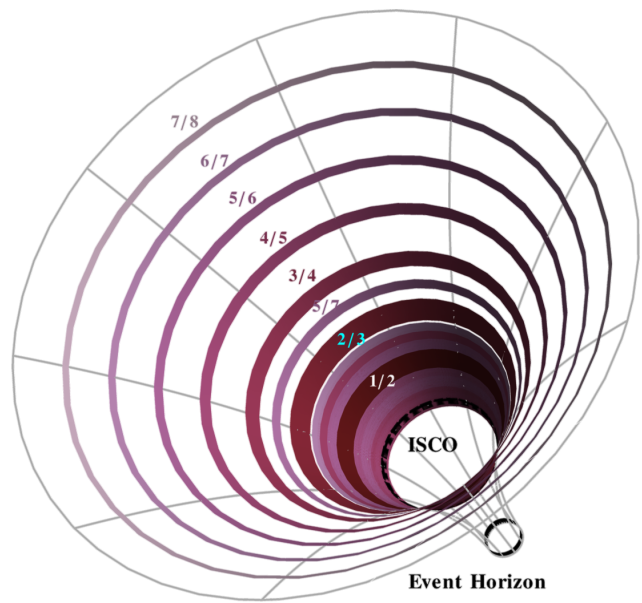}
\caption{
The location of low order resonances around a black hole \cite{ourPRL}. Here low order resonances
are plotted superimposed on a embedding diagram to give an indication of how strongly the spacetime in their vicinity is curved. The line-width in each case is scaled with $3/O_\kappa$ where $O_\kappa = m+n$ to give an indication of the relative importance of a particular radius. This image gives an over estimate of the importance of higher order resonances since the correct scaling according to Eq.~\eqref{eqKAM} is $K/O_\kappa^3$. 
\label{SaturnsRings}}
\end{figure}
Table~\ref{TabRes} indicates the characteristic length and timescales associated with  resonances in dimensionless units and as well as in units characteristic of the region around Sgr A*.  In Fig.~\ref{SaturnsRings} we have plotted all the low order resonances with order $O_\kappa=n+m\leq 15$ superimposed on an embedding diagram for a non-spinning black hole. This plot gives an indication of how strongly the spacetime for a particular resonant surface is curved and the relative extent of the regions of influence of the low-order resonances.
Note the accumulation of resonances near the ISO which can also be observed in Fig. ~\ref{fig:spine0} and persists for the spinning case. In Fig.~\ref{SaturnsRings} we have scaled the width of the lines demarcating each resonance by $K(\epsilon)/O_\kappa$ with an arbitrary choice of $K=3$  to result in good rendering.  Recall that Arnold's criterion, in Eq. \eqref{eqKAM},  governing the persistence of tori  scales as $K/O_\kappa^3$. Higher order resonances are thus likely to be strongly suppressed compared to the schematic representation given here.

The three lowest order resonant tori, $1/2, \  1/3, \ 2/3$ whose parameters are indicated in bold in Table~\ref{TabRes} are mathematically most likely to break.  Of these it is likely that astrophysically the $2/3$ resonance will have the greatest probability of being directly observable based on the following arguments: (i) For an EMRI, the impact of the broken resonance is likely to be larger if the perturbing object remains in the vicinity of the resonance for a considerable amount of time. The $1/2$ and $1/3$ resonances lie at $4 R_S$ and $3.4R_S$ respectively, close the ISO at $3R_S$ where the radiation reaction force is large and the orbit is transitioning to the plunge phase.
The time and number of orbits that an in-spiralling object of mass $\mu$ spends near a given radius roughly scale as $t\sim p^4/(\mu/M)$ and  $N\sim p^{5/2}/(\mu/M)$, thus the influence of the $2/3$ resonance can accumulate longer than for the $1/2$ and $1/3$ resonances. See also Ref. \cite{2014PhRvD..89h4036R} for more precise estimates.
(ii) For electromagnetic observations, the fact that the $1/2$ and $1/3$ resonances lie in the region of high curvature and possibly near or on the edge of the accretion 
disc is likely to confuse any distinct signal originating from this region. The $2/3$ resonance on the
other hand is located at roughly $5.4 R_S$ further out  of the potential well, making it easier for electromagnetic radiation to escape.
(iii) The final reason for expecting the $2/3$ resonance to dominate is akin to the discussion in Sec.~\ref{Sec:ROT} where it was argued that quadrupolar gravitational potential perturbations will preferentially excite the $2/3$ rather than lower order resonances.

If the resonance condition is
satisfied the particle motion either passes through the resonance, or, when the resonance dominates over dissipation so that $F_\bfk >F_\bfzero$ in Eq.\eqref{eq:FFourier} holds and depending on the initial conditions, the object can temporarily
be captured in the resonance. For captured particles the resonance effectively fixes the orbital frequency ratio of the object's orbit which manifests as a plateau in its rotation curve, as seen in Fig.~\ref{rotationNo}.  This is indicative of a stable, attracting resonance. 
    Repelling resonances, typified by an inflection point or jump in the rotation curve, are associated with unstable periodic orbits \cite{ Whiteman1977,Contopoulosbook} and will be short-lived in the presence of dissipative effects. The detailed study of the dynamics when a small mass enters the resonance region under various forms of perturbation is the subject of future work 
and has been studied in a particular Newtonian case in \cite{Chicone2}.   
In the event of resonant capture the orbital parameters will linger  within the resonance surface, possibly altering constants of motion  because of the interchange of 
energy and angular momentum between the perturbation and the orbit. 
When the gravitational radiation reaction is significant, the dissipation will most likely cause the  orbit to evolve towards a lower energy state, i.e. increasingly circular equatorial configurations, 
corresponding to the lower right-hand corner of Figs. \ref{eplELZ} and \ref{eplELZ2}.

Now consider a collection of  particles in a low-density accretion disk around an astrophysical black-hole. As explained in the introduction, their orbits will be influenced by a number of small perturbations which could cause them to become captured by a resonance, since the gravitational dissipation is small for low mass ratios. The evolution towards prograde circular equatorial orbits during resonance will limit the collisional interaction of the swarm of trapped particles, potentially creating a cohesive structure.
It is thus possible that the characteristic structure associated with the resonances, such as the resonance zones visualized in Fig. \ref{SaturnsRings}, will be imprinted on any thin disk surrounding a black-hole in the form of density inhomogeneities in much the same way as the resonant structures imprinted on Saturn's rings \cite{rings}.
Unlike the rings of Saturn where matter largely remains captured indefinitely, the black-hole rings will be dynamical because radiation reaction will alter the resonant structures and enable escape from resonance (as can be quantified using e.g. the methods in \cite{Robinson}).
In this scenario,  when a trapped over-density becomes massive enough for the radiation reaction force to become important the black-hole ring will partially disintegrate, depositing an over-density of matter on the next inward ring in the disk. There, the  matter will again be trapped for some time before continuing the in-fall. Thus in one related catastrophic event the whole ring structure will re-adjust, with the emitted radiation carrying an imprint of the particular resonances involved. Provided the accretion rate is slow enough
the ring structure will reform after each radiation reaction dominated ring collapse event. Since the radiation reaction force scales with mass ratio it is further expected that there will be a segregation in the particle sizes found in each ring. The tendency for captured orbits to evolve to a more circular, equatorial configuration is expected to minimize the ejection of simultaneously captured particles due to collisional interactions.

Another mechanism that could excite the resonant ring structure and lead to a distinct signature in the emitted radiation is the collision of a compact object with the matter in successive rings. The resulting collisional hot spot of excited gas will rotate with an azimuthal frequency set by the characteristic resonance surface. A detailed study of the possibility of a ring structure, its dynamics and observational signatures is left for future work. 

We now conduct a very precursory search for phenomena that could possibly be associated with the resonant structure around black holes. The closest super-massive black hole at the center of our galaxy presents an interesting test-bed.
Recent monitoring of Sgr A* with the 1.3mm VLBI showed time-variable
structures on scales of $\sim 4 \, R_s$ \cite{2008Natur.455...78D, 2011ApJ...727L..36F}.
The physical origin of this structure is not yet clear, but it is interesting to note that the scale is similar to
that of the low-order resonances given in Table I. As discussed above
it is possible that temporary capture
of material or gas near  the resonance location could lead to a time-varying signature in
the photon emission. For argument's sake, suppose that the origin of the structure at $\sim 4 R_S=8M$  is in fact the $2/3$ resonance but displaced from the non-spinning value listed in Table~\ref{TabRes} because Sgr~A* has spin. Using Eq. \eqref{spinEQ} the spin displacement of the prograde 2/3 resonant surface is $p_+=10.8 - 5.36 a$.  From the amount of spin displacement of the resonance needed to match the observed structure one could then infer that Sgr A* has $a=0.5$. The plausibility of identifying this structure with the 2/3 resonance could be confirmed if the variability has characteristic timescales of slightly less than an hour. The increase in sensitivity of the VLBI measurements will enable resolving more of the horizon scales, and it will be fascinating to see if the resonance structure can be revealed.  Table~\ref{TabRes} provides a quick reference for the possible characteristic time and length-scales.  Note that because the coefficients in Eq. \eqref{spinEQ} differ for different resonances, 
observing more than one 
resonant
structure places an independent check on the veracity of this method of determining the spin, since the displacement of both resonances due to spin must be consistent.

Another example of a phenomenon that could potentially be associated with the orbital resonances is
the quasi-periodic oscillations (QPOs) observed in the X-ray spectra of several black
hole candidates. Four stellar mass black hole systems exhibit quasi-periodic variability
with peaks at harmonic pairs of frequencies in a 2/3 ratio, one also shows an additional 3/5 ratio
\cite{2006ARA&A..44...49R}. Recently, QPOs have also been identified in a super-massive
black hole \cite{2008Natur.455..369G} and in a tidal disruption event
\cite{2012Sci...337..949R}. A definitive physical explanation for the origin of the QPOs
is currently lacking. Numerous models have been proposed, including orbital resonances of any
combination of the three orbital frequencies and their corresponding beats
\cite{2004ApJ...606.1098S,2001A&A...374L..19A}, accretion disk oscillations with
nonlinear effects \cite{2001ApJ...559L..25W, 2011ApJ...726...11J}, or variations in the geometry of the
accretion flow \cite{2009MNRAS.393..979L}. 
For the case where both the $2/3$ and $3/5$ ratios are observed, it is very interesting to note from Table~\ref{TabRes} that these two resonances are in fact nearest neighbors in $p$, with the $3/5$ occurring at just slightly smaller $p$ values than the $2/3$ resonance. A disruption event at the $2/3$ resonance could excite an event at the $3/5$ resonance because of their physical proximity.  Measuring the  frequencies  of the observed resonances in QPO's will give further clues as to whether they can correctly be attributed to orbital resonances around Kerr or whether other physics dominate over the orbital dynamics.

The concentration of low order resonances near the black hole and their absence further out has important implications for the key science objective of testing the no-hair theorems using a super-massive black hole such Sgr A*.
The no-hair theorems state that, provided the cosmic censorship and causality axioms hold, if the black hole's mass and spin are known the quadrupole moment is fixed.
Liu et al. \cite{Liuetal12} have
shown that recording the time of arrival 
signals of a pulsar orbiting Sgr A*, with orbital period $\sim 4 $ months,  for several
years with the Square Kilometer array (SKA)
will enable us to measure the mass of Sgr A* to a precision of $10^{-6}$,  the spin $10^{-3}$ and the quadrupole moment to $10^{-2}$. This will allow a definitive test of the
no-hair theorems. The detection of a pulsar even closer to the central object could allow the extraction of additional multipole moments through long term monitoring, thus mapping out more and more details of the structure of the central black hole.

The analysis in this paper assures us that orbits around Sgr A* with a period of the 
order of weeks to months are sufficiently far from the low-order resonances  that the 
KAM theorem guarantees  quasi-periodic motion and the persistence of invariant tori  under perturbations. 
This result implies that frequency drifts computed using perturbative methods based on averaging, as done in \cite{2011CQGra..28v5029S}, accurately describe the physical system that is effectively free of stochastic motion. 
It also ensures that tracking the regular motion of a pulsar will reflect a map of the gravitational potential it samples. 
We conclude this section by mentioning in passing the relevance to future gravitational wave detectors such as eLISA and their potential to directly probe resonant dynamics.
A detector  sensitive to the frequency band 
$\sim 10^{-4}-10^{-1}$ Hz observing an EMRI near SgrA* \mbox{$M_{SgrA*} \approx 4 \times 10^6 M_\odot$}
\cite{Gillessen2008},  can probe mean radial distances ranging from the event horizon to $\sim 50M$ \cite{MGeyerThesis}. This overlaps with the region where the strongest resonances occur (see Table \ref{TabRes} ).
The potential direct detection of gravitational wave emission from a resonance transit is an exciting possibility. It does however underscore the necessity to carefully model and incorporate resonant effects in the search templates.  The loss of phase coherence as the small mass object passes through a resonance could potentially make parameter estimation difficult.

\section{Summary and Discussion}
In this paper we have computed the location and associated timescales of resonant orbits in the Kerr spacetime and comment on the observational and 
mathematical implications. We have considered 
resonances between the two fundamental frequencies corresponding to the radial 
and longitudinal motion, leaving the study of the effect of the rotational 
azimuthal frequency to future work. Our results provide a complete survey 
of the parameter space of resonant orbits, together
with simple expressions for locating the resonances valid in the strong field region. If resonant phenomena are observed these expressions could provide and easy way of determining the spin of the central black hole. We considered several examples of electromagnetic observations at different wavelengths that could be related to the capture and escape of material from resonances.

We have computed the resonant surfaces both in terms of generalized Keplerian variables 
(related to the orbital geometry)
as well as projected onto the $E$ and $L_z$ parameter space (related to the spacetime
 symmetries) to help identify promising choices of parameters for numerical 
investigations of torus breakdown in resonance regions.
We have further found an analytic expression for the rotation curve associated with
the Kerr metric as a function of $E$ and $L_z$ that can be used for comparing
to numerically computed rotation curves associated with Poincar\'{e} maps. 
We expect low order resonances such as the $2/3$ and $1/2$ resonances to have 
the strongest observable effects on orbital motion, electromagnetic emission and 
the phase of emitted gravitational waves. These resonances occur in the strong field 
regions of the spacetime at a distance of $\lesssim 5.4 \, R_s$
from the black hole and are fairly widely spaced by $\sim 1.4 \, R_s$ in the 
limiting case of a non-spinning black hole.

According to the KAM theorem low order resonances indicate where in a dynamical 
system the transition to chaos is likely to occur first.  Such a transition
requires a sufficiently large perturbation that could arise from various sources 
such as the non-Kerr nature of the spacetime, the internal structure of the probe, 
alternative theories of gravity or the presence of other bodies or gas. 
We summarized the KAM estimate and additional arguments by Arnold and Chirikov to assess
which resonant tori are expected to survive under the perturbation and where dramatic changes in the dynamics could occur.
 Combining these general bounds with our results for the resonances indicates that there is a large region
 for mean radii $ 50 \, R_s\lesssim p \lesssim 1000 \, R_s$ where resonance effects are negligible but where
 the spacetime curvature is sufficiently high that multipoles of the central object have an observable effect
 on the motion. This will enable tests of the no-hair theorem unimpeded by drastic
 changes in the dynamics.

More stringent bounds than discussed in this paper on the occurrence of strong resonances
and the transition to stochastic behavior require detailed studies of the dynamics near 
resonant tori for the different kinds of perturbations. We noted in Sec. \ref{Breakdown} 
that for EMRIs the gravitational radiation reaction force, which is important in the 
region where low-order resonances occur, sets an approximate scale for the strength of 
the perturbation required to destroy the tori. In future work we intend to study the 
details of the breakup of the tori and quantify the strength of the perturbation required 
for observable consequences in each case.

Regions where lower order resonances occur also earmark the most likely positions in
phase space where averaging methods 
must be modified to account for the resonances.  
If an orbit passes through a resonance it effectively acquires a sudden change
in the frequencies whose magnitude depends on the initial conditions.
For sufficiently strong resonances, orbits can enter a
resonance region, linger near it and subsequently escape. This manifests as a plateau in the
frequency evolution similar to that seen in the rotation curve of Fig. \ref{rotationNo} but
tilted because of radiation reaction. More work will be needed to determine how well the
frequencies must be resolved to detect such a plateau and exploit the measurements to
determine the system parameters with LISA.

In addition to exploring the parameter space of resonances in this paper, we characterize in the Appendix the locations of
the innermost stable orbit beyond which geodesics plunge into the black hole. This
provides a useful benchmark for the resonance locations and also indicates the region
where zoom-whirl behavior would occur and where higher order resonances accumulate.

\section{Acknowledgments} 
This work was supported in part by National Science Foundation Grants PHY-0903631 and PHY-1208881, the Maryland Center for Fundamental Physics, the Square Kilometer Array project in South Africa and the National Institute of Theoretical Physics in South Africa. 

\appendix

\section{Location of the Innermost Stable Orbit}
\label{ISOapp}

\begin{figure*}
\includegraphics[width = \textwidth]{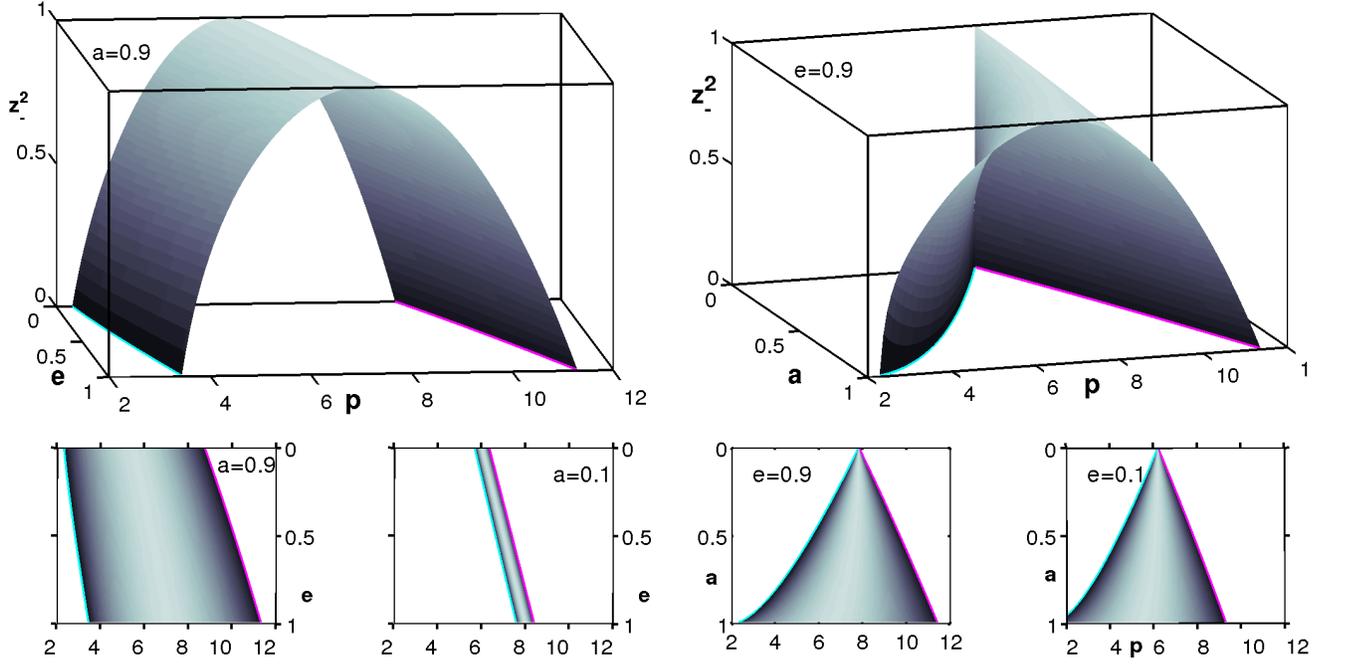}
\caption{Graphical representation of ISO surface constructed using Eq's. \eqref{ISOzm}\eqref{r4roots}. 
The ISO surfaces displayed in the top two images shows the same qualitative $'U'-'V'-'I'$ transitions as the resonant orbits. Smaller images in the bottom row give the top view of the resonant surfaces for various spins and eccentricities.
The top left and the two images on the bottom left display the eccentricity dependence for different spin values. 
The ISO is linearly dependent on eccentricity (as opposed to the quadratic dependence observed in resonant surfaces). The spin dependence at various fixed eccentricities is illustrated in the right three plots where the typical $'V'$ profile is obvious irrespective of eccentricity.  
\label{fig:ISO}}
\end{figure*}
In this appendix we explore the location of the innermost stable orbit (ISO) using the notation and variables given in Sec. \ref{EOM}. Observationally this is an important location in the black hole spacetime since it demarcates the radius from which the test mass will enter the plunge phase of its orbit. Near the ISO a non-circular orbit exhibits zoom-whirl behavior as the radial frequency goes to zero while $\omega_\theta$ and $\omega_\phi$ remain finite. The test mass will linger near its periastron for many periods of the $\phi$ and $\theta$ motion, then rapidly zoom out to apastron and back, giving rise to a characteristic signature in the emitted gravitational waves.
 When tracking the frequency evolution of a particular orbit, this determines the final set of $\omega_\phi$, $\omega_{\theta}$ frequencies that will be recorded before the test mass begins its plunge into the black hole.
 When discussing resonant orbits of a spinning black hole it is often useful to view their location relative to the location of the ISO for the same spin, inclination and eccentricity.  The ISO surface (where $\omega_r$ vanishes) shares many of the qualitative features of the resonance surfaces (where the combination of frequencies $m\omega_r-n \omega_\theta$ vanishes) discussed in the main body of the text and both are related to degenerate structures in the phase space that are broken under small perturbations.

The ISO occurs when the potential barrier in the radial potential no longer shields the orbit from the singularity. Mathematically the condition that defines this orbit is that the middle two roots of Eq.\eqref{Rpot} coincide,
\begin{align}
r_3=r_2 = \frac{p}{1+e}.
\end{align}
This allows us to express the $\varpi_+$ and $\varpi_{\times}$ variables as.
\begin{align}
\varpi_{+} &= r_4 + \frac{p}{1+e}, & \varpi_{\times}= \frac{r_4 \ p}{1+e} .\label{vpr4}
\end{align}
Eq. \eqref{wpcond} allows us to find an expression for
$z_-^2$ in terms of $r_4$, 
\begin{align}z_{-}^2&=
\frac{p \left(\tilde{r_4}(e-3)      -(e+3) p \right)}{2 a^2 (e-1) (e+1)^2} \notag\\
&+ \frac{p\sqrt{\left(e-3\right)^2 \tilde{r_4}^2+2
   \left(e^2+7\right) p \tilde{r_4}+(e+3)^2 p^2}
}{2 a^2 (e-1) (e+1)^2} \label{ISOzm}
,\end{align}
where we have set  $\tilde{r_4} = (1+e)r_4$ to yield a more compact expression.
Eq. \eqref{wptquad} provides a quadratic equation for $r_4$ that allows us to determine its value  in terms of $p$ and $e$.
The solution can be written down in the form,
\begin{align}r_4=(-B_{r_4} + \sqrt{\Delta_{r_4}})/A_{r_4} ,\label{r4roots}
\end{align}
 with
\begin{widetext}
\begin{align}
\Delta_{r_4}&= 64 (e+1) p^3
\left[p-(e-1) \left(\tilde{a}-1\right)\right]
\left[p+(e-1)   \left(\tilde{a}+1\right)\right]
\left[p-(e+1)   \left(\tilde{a}+1\right)\right]^2
\left[p+(e+1)   \left(\tilde{a}-1\right)\right]^2, \notag\\
A_{r_4}&=
a^4 (e-1)^2 (e+1)^4-4 a^2 (1-e)(3-e) (e+1)^3 p+2 (e+1)^2 p^2 \left(a^2 (3+e)(e-1)+2
   (e-3)^2\right)
\notag\\
&-4
   (e^2+7)( e+1) p^3+(e+3)^2 p^4, \notag\\
B_{r_4} &= a^4 (e-3) (e-1) (e+1)^3 p
-4 a^2 (e+1)^2 \left(e^2-2 e+3\right) p^2
+2 (e+1) p^3 \left(a^2 (e^2-5) +2
   e^2+14\right)\notag\\
&-4 \left(e^2+2
   e+3\right) p^4
+(e+3) p^5.
\end{align}
\end{widetext}
We have set $\tilde{a}=\sqrt{1-a^2}$.
In Eq. \eqref{r4roots} we chose the `+' root of the quadratic to yield the correct behavior for $z_-^2$.
The ISO surface exhibits many of the characteristics seen in the resonance surfaces studied in Sec.~\ref{SolSec}. These include the fact that
as inclination increases the $p$ values of the prograde/retrograde ISO's respectively increase/decrease until the two branches coincide for \mbox{$z_-^2 = 1$}.
This behavior is due to the fact that polar orbits are less influenced by the spin of the black hole. Also note that for \mbox{$z_-^2=1$},  $dz_-^2/dp =0$.  A graphical representation of the ISO surface is given in Fig. \ref{fig:ISO}

We now give a number of easily evaluated formulae for special parameter values.
For circular orbits around a maximally spinning black hole ($e=0$, $a=1$), the behavior of $z_-^2$ as a function of $p$ is described by,
\begin{align}
z_1^2=-\frac{p^2 \left(-3 \sqrt{p}+2 \sqrt{3 p+2 \sqrt{p}+3}-3\right)}{3
   \sqrt{p}-1}.\end{align}
Furthermore when $e=1$, $a=1$,
\begin{align}
z_1^2=\frac{8 \sqrt{2} p^{7/2}-3 p^4-4 p^3+4 p^2}{4 \left(9 p^2+4 p+4\right)}.
\end{align}
On the equatorial plane $z_-=0$ and $r_4=0$. Just as for the resonances the strongest spin dependent effects can be observed here. The behavior of the ISO
can be well characterized by merely examining the behavior on the equator.
As inclination increases, the location on the ISO will lie between the extremes of the prograde and retrograde values found on the equatorial plane.  The characteristic $'V'$- $'I'$ transition seen in the resonances occurs for the ISO as well. If $r_4=0$ only the constant term (with respect to $r_4$) in Eq. \eqref{wptquad} with Eq. \eqref{vpr4} substituted  remains. This term results in a  quartic equation for $p$ that implicitly defines the location of the ISO on the equator, as also found in  \cite{2003PhRvD..67d4004O},
\begin{align}
 (p-6-2e)^2  p^2
+a^4(1+e)^2(3-e)^2 \notag\\
 -2a^2(1+e)p \left[ 2 (e^2+7)
 +
(3-e)p\right]
 =0\label{ISOEQ}
.\end{align}
If  the black hole is non-spinning, $a=0$, then  $p=6+2e$. If it is maximally spinning, $a=1$\cM, there are 3 roots to Eq. \eqref{ISOEQ}, namely
$p=5 +e \pm 4\sqrt{1+e}$  and a double root at $p=1+e$.
Of these the correct limiting cases for the retro and prograde ISO orbits are $p_-=5 +e + 4\sqrt{1+e}$  and $p_+=1+e$ respectively.
Thus the maximum splitting between the pro and retrograde branches on the equatorial plane is $p_--p_+ = 4(1+\sqrt{1+e})$.

 If  $a\neq 0$ the ISO can be found by solving the quartic. (For plotting purposes it is easier to consider Eq. \eqref{ISOEQ} to be a  quadratic equation for $a^2$ and plot the square root of the result as a function of $p$.)
For convenience, the leading order spin dependence is also given here,
\begin{align}
\label{ISOpeqa0}
p_{\mp} = 6+2 e \pm 4a \sqrt{2 \left(\frac{1+e}{3+e} \right) }+ O(a^2)
\cM{.}\end{align}

\section{Properties of Carlson's integrals}
\label{app:Carlson}

Carlson \cite{1995NuAlg..10...13C}   \cite{NIST:DLMF}  \cite{Olver:2010:NHMF}  has provided us with a number of symmetric, rapidly convergent schemes to evaluate elliptic integrals both numerically and analytically.
This appendix summarizes the identities and properties associated with Carlson's integrals 
 used in the body of the paper.
Carlson showed that any elliptic integrals of the form
\begin{align}
I_1 = \int_y^x \frac{dt}{\sqrt{(\alpha_1+\beta_1t)(\alpha_2+\beta_2t)(\alpha_3+\beta_3 t) (\alpha_4 +\beta_4 t)}}
\label{I1start}
\end{align}
can be expressed as
\begin{align}
I_1 = 2 R_F(U_{12}^2,U_{13}^2, U_{14}^2) \label{CI1}
,\end{align}
where
\begin{align}
U_{ij} = (X_i X_j Y_k Y_m + Y_i Y_j X_k X_m)/(x-y), \notag\\
X_i = (\alpha_i+\beta_i x)^{1/2},\ \ \  Y_i=(\alpha_i+\beta_iy)^{1/2} \notag\\
\end{align}
and
\begin{align}
R_F(\alpha,\beta,\gamma) &= \frac{1}{2} \int_0^\infty \frac{dt}{\sqrt{(t+\alpha)(t+\beta)(t+\gamma)}} \label{rfdef}
.\end{align}
This symmetric representation greatly reduces the number of parameters  from
10 in Eq. \eqref{I1start} to the three arguments in Eq. \eqref{rfdef}.
All integrals have the same boundary conditions and the arguments enter only in the denominator of the integrad. An added advantage of using Carlson's integrals is that they obey a number of identities that make manipulation of the parameters that enter as arguments possible without necessarily evaluating the integral. These identities include the duplication theorem,
\begin{align}
R_F(\alpha,\beta,\gamma) = 2R_F(\alpha+\lambda, \beta+\lambda, \gamma+\lambda),
\end{align}
where $\lambda = (\alpha\beta)^{1/2}+(\alpha\gamma)^{1/2}+(\beta\gamma)^{1/2}$;
the fact that Carlson's function is homogeneous of degree $-\frac{1}{2}$,
\begin{align}
R_F(\lambda \alpha,\ \lambda \beta, \ \lambda \gamma) = \lambda^{-1/2} R_F(\alpha,\beta,\gamma), \label{HomoGen}
\end{align}
for any $\lambda$ and a number of special symmetric cases that can  easily be  evaluated,  \begin{align}
R_F(\beta,\beta,\beta) &= \beta^{-1/2}, &R_F(0,\beta,\beta) &= \frac{\pi}{2} \beta^{-1/2}.
\label{Rsymm}
\end{align}
If the first argument is $\alpha=0$ a restricted version of the duplication theorem also holds:
for $x>0$, $y>0$, $z=(x+y)/2$,
\begin{align}
R_F(0,x^2,y^2)= R_F(0,xy,z^2).
\label{Rfxysq}
\end{align}
For small deviations from the symmetric case it is possible to construct a rapidly converging series. We now derive this series for $R_F(0,y+\delta,y-\delta)$.
Start with the integral representation of $R_F$, Eq.  \eqref{rfdef}
in the special case,
\begin{align}
R_F(0,y+\delta,y-\delta)=\frac{1}{2}\int^\infty_0 \frac{dt}{\sqrt{t(t+y+\delta)(t+y-\delta)}}
,\end{align}
and make a Taylor series expansion of the integrad in terms of $\delta$. Integrate the result term by term to yield,
\begin{align}
\label{taylor}
R_F(0,y+\delta,y-\delta)=\frac{\pi }{2 \sqrt{y}} \left(1
+\frac{3 \delta ^2}{16 y^2}
+ \frac{105
   \delta ^4}{1024 y^4}\right)\notag\\ +
\frac{\pi }{2 \sqrt{y}} \left(
\frac{1155 \delta ^6}{16384 y^6}+
 \frac{225225 \delta ^8}{4194304 y^8}  + O(\delta^{10}) \right).
\end{align}
Since only quadratic terms in the parameter $\delta/y$ appear this series converges very rapidly.
To complete the discussion on Carlson's integrals and mainly for comparison with other work we now give the relationship between Carlson's integrals and some of the elliptic functions,
\begin{align}
R_f(0,1-k^2,1) &= K(k),\\
R_F(0,\alpha,\beta)&= 
\beta^{-1/2}K(\sqrt{1-\alpha/\beta}). \label{RFellip}
\end{align}
Here $K(k)$ is the complete elliptic integral of the first kind.

\bibliography{../BholesNemadon,../ResonanceRefs}

\end{document}